\documentclass[twocolumn,longauthor,tighten,astrosymb]{aastex631}

\usepackage{graphicx}
\let\tablenum\relax
\usepackage{siunitx}

\received{April 2, 2024}
\revised{June 26, 2024}
\accepted{June 29, 2024}

\shorttitle{Resolving Twin Jets and Twin Disks with JWST and ALMA in WL 20}
\shortauthors{Barsony et al.}

\begin{document}

\title{Resolving Twin Jets and Twin Disks with JWST and ALMA:  \\
The Young WL 20 Multiple System}

\correspondingauthor{Mary Barsony}
\email{marybarsony60@gmail.com}

\author[0009-0003-2041-7911]{Mary Barsony}
\affiliation{13115 Dupont Road, \\
Sebastopol, CA 95472, USA}

\author[0000-0001-5644-8830]{Michael E. Ressler}
\affiliation{Jet Propulsion Laboratory, California Institute of Technology, \\
4800 Oak Grove Drive, \\
Pasadena, CA 91109, USA}

\author[0000-0002-5714-799X]{Valentin J.M. Le Gouellec}
\altaffiliation{NASA Postdoctoral Fellow}
\affiliation{NASA Ames Research Center,\\
Space Science and Astrobiology Division, \\
M.S. 245-6, \\
Moffett Field, CA  94035, USA}

\author[0000-0002-9470-2358]{{\L}ukasz Tychoniec}
\affiliation{Leiden Observatory, \\
Leiden University, \\
P.O. Box 9513,\\
2300RA Leiden, The Netherlands}

\author[0000-0002-6312-8525]{Martijn L. van Gelder}
\affiliation{Leiden Observatory, \\
Leiden University, \\
P.O. Box 9513, \\
2300RA Leiden, The Netherlands}



\begin{abstract}

We report the discovery of jets emanating from pre-main-sequence objects exclusively at mid-infrared wavelengths,
enabled by the superb sensitivity of JWST's Mid-InfraRed Medium-Resolution Spectrometer (MIRI MRS) instrument. These jets are observed only in
lines of [NiII], [FeII], [ArII], and [NeII]. The H$_2$ emission, imaged in eight distinct transitions, has a completely 
different morphology, exhibiting a wide-angled, biconical shape, symmetrically distributed about the 
jet axes.  Synergistic high-resolution Atacama Large Millimeter/submillimeter Array (ALMA) observations
resolve a pair of side-by-side edge-on accretion disks lying at the origin of the twin mid-infrared jets.
Assuming coevality of the components of the young multiple system under investigation,
the system age is at least (2 $-$ 2.5) $\times$ 10$^6$ yr, despite the discrepantly
younger age inferred from the spectral energy distribution of the combined edge-on disk sources.
The later system evolutionary stage is corroborated by ALMA observations of CO(2$-$1), $^{13}$CO(2$-1$), and 
C$^{18}$O(2$-$1), which show no traces of {\it molecular} outflows or remnant cavity walls.
Consequently, the observed H$_2$ structures must have their origins in wide-angled disk winds,
in the absence of any ambient, swept-up gas. 
In the context of recent studies of protostars, we propose an outflow evolutionary scenario
in which the molecular gas component dominates in the youngest sources,
whereas the fast, ionized jets dominate in the oldest sources, as is the case for the twin jets discovered in the WL 20 system.
\end{abstract}



\section{Introduction}\label{sec:intro}
Most stars form in multiple systems \citep{Offner2023}, defying simple theoretical collapse models \citep{Shu1977}.
It is now recognized that in addition to gravity and thermal pressure, turbulence, magnetic fields, and
interactions between the members of multiple systems all play a role in their formation \citep{Murillo2016, Tychoniecetal2024}.
Observational studies of multiple formation are, therefore, crucial for further progress.
How do the individual members of a multiple system coevolve? 

The WL 20 triple system in Ophiuchus (Oph) is of great interest in this context, since it is a member of a rare class of multiple systems known as InfraRed Companion (IRC) systems, 
in which one member has the appearance of being significantly younger than its companions, despite the fact that they have all formed from a single core {\it e.g.,}
T Tau, Glass I, Haro 6-10, Z CMa, XZ Tau, DoAr24E
\citep{Dycketal1982, Chelli1988, LeinertHaas1989, Ghez1991, Koresko1991, Haas1990, Barsonyetal2001}.
Such systems are particularly interesting for pre-main-sequence evolutionary studies, since they pose
a puzzle as to why one of two or more coeval components appears significantly redder, and, in this case,
more luminous than its companions.

WL 20 was discovered in the course of a 2 $\mu$m bolometer survey, carried out with a 12$^{\prime\prime}$ beam,
of the 10.5$^{\prime} \times 10.5^{\prime}$ region exhibiting the the strongest C$^{18}$O emission in the Rho Ophiuchi ($\rho$ Oph) cloud 
\citep{Wilking1983}. When near-infrared arrays became available, WL 20 was resolved first into a binary
and, eventually, into a triple system, at 2.2 $\mu$m  \citep{Rieke1989, Barsony1989, StromKepnerStrom1995}.
WL 20E ($K=10.13$) and WL 20W ($K=10.40$) have an angular separation of 3.17$^{\prime\prime}$ at P.A. 270$^\circ$ E of N,
whereas the IRC, WL 20S, lies 2.26$^{\prime\prime}$ from its nearest neighbor, WL 20W, at a P.A. of 173$^{\circ}$ \citep{ResslerBarsony2001}.

The first near-infrared spectra of the two brighter components, obtained with $R \equiv \lambda/{\delta\lambda} \le 1000$, 
determined spectral types consistent with K$-$M for WL 20E through an extinction of
$A_V=15.4$ and K7$-$M0 through $A_V=18.1$ for WL 20W \citep{GreeneMeyer1995}. Subsequent higher-resolution ($R \sim 1200$) near-infrared
spectra established spectral types of K6 for WL 20E (GY 240B) and M0 for WL 20W (GY 240A), both seen through $A_V = 16.3$ \citep{LuhmanRieke1999}. 
It took NIRSPEC on the Keck II 10.4-meter telescope to finally obtain a spectrum of the fainter ($K=12.6$)
IRC, WL 20S. The $R = 2200$ spectra of each component refined the spectral types of WL 20E to K7 IV/V and WL 20W to M0 IV/V,
using veiling independent line ratios, whereas the spectrum of WL 20S is so heavily veiled that no absorption lines could be detected in its spectrum. 
Nevertheless, the spectral shape and $\delta K = 2.2$ mag brightness difference between WL 20S
and WL 20W constrains WL 20S to have an infrared excess $r_K < 0.9$, but with an {\it additional} $A_V=25$, relative to its neighbors \citep{BGB2002}.

Mid-infrared observations of WL 20 were first acquired with a 6$^{\prime\prime}$ (or 8$^{\prime\prime}$)
aperture at 10 $\mu$m, which did not resolve the multiple system \citep{LadaWilking1984}. 
In the morphological classification scheme of pre-main-sequence spectral energy
distributions (SEDs) devised by Lada and coworkers (e.g., \citet{Lada1987, ALS1987}),
WL 20 was classified as a Class I object \citep{WLY1989}. 
Diffraction-limited mid-infrared imaging on the Keck II telescope allowed 
spatially resolved determination of each component's SED, confirming the previous Class II
spectroscopic classifications of WL 20E and WL 20W \citep{GreeneMeyer1995}, but demonstrating the 
Class I SED of WL 20S \citep{ResslerBarsony2001}. 

The first millimeter continuum detection of WL 20 was with the IRAM 30-meter telescope at 1.3 mm
with an 11$^{\prime\prime}$ beam \citep{AndreMontmerle1994, MAN1998}. Interferometric observations
with the six-element Owens Valley Radio Observatory (OVRO) array were required to identify WL 20S as the
source of the millimeter dust continuum emission associated with the system \citep{BGB2002}.
The single-telescope and interferometric flux measurements were consistent, implying a compact source
structure origin, with no emission from any envelope component.  
This conclusion is further corroborated
by comparison of HCO$+\ J\ =\ 4 \rightarrow 3$ and 850 $\mu$m dust maps of WL 20S: Since a hallmark of the earlier
evolutionary stage is the presence of a centrally condensed envelope, such sources should exhibit 
HCO$+\ J\ =\ 4 \rightarrow 3$ emission coincident with the dust continuum peak, since
this transition has a high critical density ($> 10^6$ cm$^{-3}$), unique to
the dense gas located in the inner regions of protostellar envelopes. Such a spatial coincidence was not
found for WL 20S, leading to the conclusion that it lacks an infall envelope component, and 
sports a Class I SED due to its edge-on orientation \citep{vanKempenetal2009}.

Intriguingly, the 12.81 [NeII] line was detected in WL 20 in a 4.7$^{\prime\prime}$ slit at R$=$600 by {\it Spitzer's IRS},
but was undetected in a much narrower, 0.4$^{\prime\prime}$ slit by the Very Large Telescope's (VLT's) VISIR at R=30,000, 
implying a spatially extended origin for the emission \citep{Saccoetal2012}. 

Thermal radio jets are typically associated with Class 0/I protostars \citep{Beltran2001,
Carrasco-Gonzalez2012, Diaz_Rodriguez2021}. A 6 cm source detected with the Very Large Array (VLA) in an 
$11^{\prime\prime} \times 5^{\prime\prime}$ beam was first associated with WL 20 in
the survey of two dense cores, A and E/F, in the $\rho$ Oph star-forming cloud \citep{Leous1991}. 
Higher-angular-resolution JVLA 3.0 cm maps (5.1$^{\prime\prime} \times 2.4^{\prime\prime}$ beam at P.A.$=\,-5^{\circ}$) 
associated the radio jet with WL 20S \citep{Rodriguez2017}.

To further delve into the mysteries posed by this system, we have acquired JWST MIRI MRS
integral field unit imaging spectroscopy encompassing all three components, 
covering the 5 $-$ 28 $\mu$m wavelength range, supplemented  by high-resolution ALMA data.

The paper is structured as follows:  Observations and Data Reduction for MIRI MRS and for the ALMA data are presented in $\S$2.1
and $\S$2.2, respectively. MIRI MRS results are provided in $\S$3.1 as follows:  $\S$3.1.1 covers continuum images,
$\S$3.1.2 shows the on-source spectra,  $\S$3.1.3 highlights jet images and spectra in both low- and high-excitation lines,
and $\S$3.1.4 features the molecular hydrogen line images.
ALMA results are presented in $\S$3.2:  $\S$ 3.2.1 features Band 4 (1.9 mm) continuum images and $\S$3.2.2 both Band 6 (1.3 mm)
continuum and CO(2$-$1), $^{13}$CO(2$-$1), and C$^{18}$O(2$-$1) line maps.
Discussion is contained in $\S$4 and Conclusions in $\S$5.

\section{Observations and Data Reduction}\label{sec:obs}
\subsection{{\it MIRI MRS}}\label{subsec:MIRIobs}

WL 20 was observed during UTC 12-13 April 2023 
as part of Guaranteed Time Observations (PID 01236; P.I. Ressler) with JWST's
MIRI \citep{rieke2015, wright2015, wright2023}
in its MRS mode \citep{wells2015, labiano2021, argyriou2023}.
Observations were acquired with a two-point dither pattern.
The pointing center for all observations was $\alpha_{2000}$ $=$ 16h 27m 15.77s, $\delta_{2000}$ $-24^{\circ}$ 38$^{\prime}$ 44.3$^{\prime\prime}$.
To maximize on-source integration time,
no dedicated background observations were undertaken. Instead,
background observations acquired earlier under PID 01236 are used to subtract the telescope
background and detector artifacts during the data processing.
All three gratings (A, B, and C) were employed through all four MIRI Channels,
using the FASTR1 read mode, thereby covering the entire 4.9 $-$ 28.6 $\mu$m spectral range at resolutions ranging from 3500 $\ge$ R $\ge$ 1500.
Integration times were 15 minutes 55 seconds through the short grating, and 15 minutes 52 seconds in each of the medium and long gratings. 
In addition, the F1500W filter was chosen for parallel off-source imaging for the duration of the MIRI MRS observations.

There are three necessary data-processing steps for producing usable science data from
the MIRI MRS raw data.
These were performed using the JWST pipeline version 1.11.4 \citep{Bushouse2023} using reference context
{\tt jwst\_1118.pmap} of the JWST Calibration Reference Data System (CRDS; \citet{Greenfield2016}).
Level 1 processing was performed with the default settings in
 the {\tt Detector1Pipeline}. 
 Level 2 processing was performed using the Spec2Pipeline.  During this step, the dedicated background of Program 01236,
 obtained at a different time and region of sky, was subtracted on the detector level in order to subtract the telescope background and detector artifacts.  
 Additionally, fringe corrections were performed using the fringe flat for extended sources (Mueller, M. et al., in prep.) 
 and detector level residual fringe corrections were applied (Kavanagh, P.  et al., in prep.). 
 Finally, the {\tt Spec3Pipeline} was run with both the outlier rejection
 and master background subtraction steps switched off. This processing resulted in 12 calibrated data cubes,
 one for each combination of channel and grating settings.

\subsection{ {\it ALMA}}

Band 6 (1.3 mm) ALMA observations of the WL 20 system were obtained as part of program 2019.1.01792.S (PI: D. Mardones). Two datasets are used, one high-angular-resolution dust continuum observation reaching $\sim\,0.11^{\prime\prime}$ with 36 sec of integration time, and a lower-angular-resolution observation reaching $\sim\,1.1^{\prime\prime}$ with 194 sec of integration time. The longer-integration time, lower-angular-resolution data target various molecular lines with high spectral resolution (i.e., 0.08 km s$^{-1}$). In this work we will focus on the CO(2$\rightarrow$1), $^{13}$CO(2$\rightarrow 1$), C$^{18}$O(2$\rightarrow 1$) transitions. We used the \texttt{tclean} task of CASA version 6.5.2 to produce the dust continuum image and molecular line channel maps, with Briggs weighting and the auto-multithresh option for the masking operations. Robust parameters of 0.0 and -0.5 were used for the dust continuum and molecular line maps, respectively. We reached noise levels of 0.27 mJy beam$^{-1}$ and 25 mJy beam$^{-1}$ per 0.125 km s$^{-1}$, for the continuum and spectral line observations, respectively,
calculated using the root-mean-square (rms) flux of an emission-free region in the image plane in each instance.

Band 4 (1.9 mm) 155 GHz ALMA observations were obtained on 2023 June 12 within program 2022.1.01734.S (PI: Ł. Tychoniec) with 30 sec integration time.
The observations were self-calibrated with the \texttt{auto$\_$selfcal} package\footnote{https://github.com/jjtobin/auto$\_$selfcal}. 
The resulting measurement set was imaged with the \texttt{tclean} procedure within CASA version 6.5.2 \citep{McMullin.Waters.ea2007}. A robust parameter of 0.5 was used, and automasking with standard parameters was applied.
The resulting image has a resolution of 0.096$^{\prime\prime}$ $\times$ 0.16$^{\prime\prime}$ and sensitivity of 0.075 mJy beam$^{-1}$ measured from the rms signal in an emission-free part of the continuum image.

\section{Results} \label{sec:results}
\subsection{\it MIRI MRS}\label{subsec:MIRI}

\subsubsection{Line-free Continuum Images:  Newly Discovered Source in WL 20S}

\begin{figure*}
\plotone{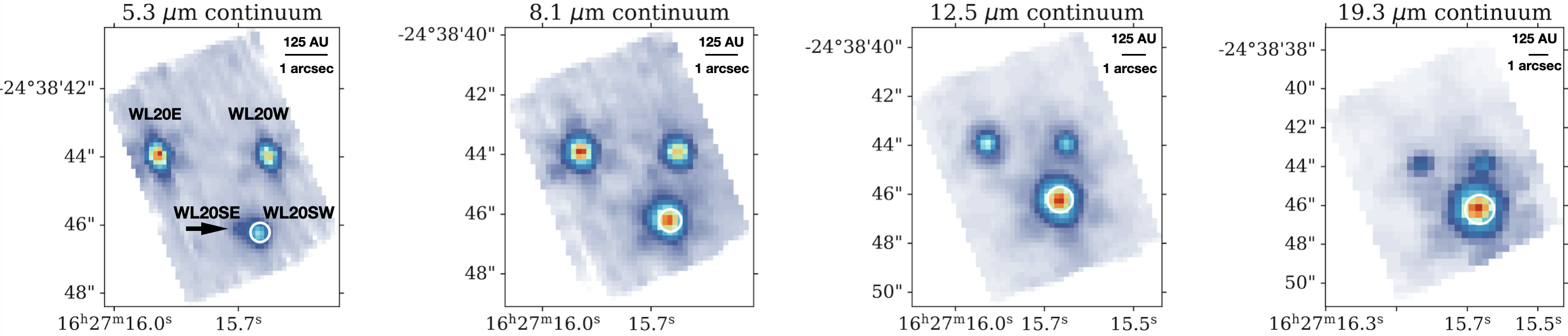}
\caption{The WL 20 system observed through the SHORT (A) subbands in each of the four MIRI MRS channels, with their varying fields of view (FOV's):
Channel 1A $-$ 4.90$\rightarrow$5.74 $\mu$m,
FOV$=3.2^{\prime\prime}\times 3.7^{\prime\prime}$; 
Channel 2A $-$ 7.51$\rightarrow$8.77 $\mu$m,  FOV$=4.0^{\prime\prime}\times 4.8^{\prime\prime}$,
 Channel 3A $-$  11.55$\rightarrow$13.47 $\mu$m, FOV$=5.5^{\prime\prime}\times 6.2^{\prime\prime}$, and
 Channel 4A $-$ 17.7$\rightarrow$20.95 $\mu$m, FOV$=6.9^{\prime\prime}\times 7.9^{\prime\prime}$.
The median wavelengths of each subband are indicated at the top of each panel.
Each component of the system is labeled in the 5.3 $\mu$m continuum image.  
The previously known InfraRed Companion source, WL 20S, is
encircled in white in each frame. The aperture encircling WL 20S scales as  2.00 $\times {\lambda \over D}$, the same aperture used to extract its spectrum shown in 
the bottom panel of Figure \ref{fig:spectra}.
The source previously known as WL 20S is now designated
as WL 20SW, to distinguish it from the fainter, neighboring source, WL 20SE, discovered in these data.  The new source
is indicated by the black arrow in the left panel, and is visible only at the shortest MIRI wavelengths,
where the spatial resolution is highest.  
}
\label{fig:WL20continuum}
\end{figure*}

Figure \ref{fig:WL20continuum} 
shows the appearance of the WL 20 system at four different line-free continuum
wavelengths through the SHORT (A) grating in each of MIRI's four MRS Channels (see Table 1 of \citet{labiano2021}).
Surprisingly, a faint, new source, WL 20SE, was discovered next to WL 20S, evident at the shorter MIRI MRS wavelengths (left panel of 
Figure \ref{fig:WL20continuum}), with a separation of 
$\sim$ 0.58$^{\prime\prime}\ \pm 0.03^{\prime\prime}$ at PA 76.1$^{\circ}\ \pm 0.5^{\circ}$ (measured E from N),   
relative to the brighter component, WL 20S. In the 8.1 $\mu$m continuum image and at longer wavelengths, however, 
this new source can no longer be separated from its brighter
companion, their fluxes being blended together due to the increasing point-spread-function (PSF) with wavelength.  
Consequently, when these two objects are spatially resolved, the IRC source, previously designated as WL 20S,
will now be referred to as WL 20SW, and its newly discovered neighbor will be called WL 20SE.

Table \ref{table:results1} lists the detected sources and source coordinates, as determined
from the WCS header coordinates produced by the JWST pipeline for MIRI MRS and refer to 
the source centroid coordinates at 5.3 $\mu$m. A second set of coordinates, derived
from the ALMA continuum observations as described in $\S$\ref{subsec:ALMA_Band4}  {\it ALMA Band 4 Data},  is also listed.

\begin{figure*}
\plotone{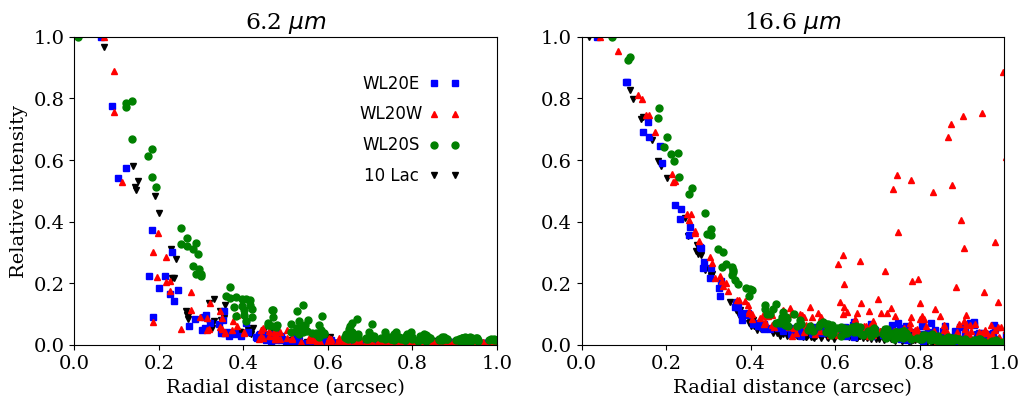}
\caption{ Left panel: 6.2 $\mu$m (data averaged over the Channel 1 Medium spectral band)
azimuthally-averaged radial profiles of the point-source
calibrator, 10 Lac (downwards-pointing, black triangles), WL 20E (blue squares),
WL 20W (upwards-pointing red triangles), and WL 20S (green circles), each
scaled to their peak pixel value.  Right panel:  Same as for the left panel, but
at 16.6 $\mu$m, data averaged over the Channel 3 - Long spectral band.
Whereas the radial profiles of WL 20E and WL 20W are consistent with
being point-sources, the radial profiles of WL 20SW clearly indicate
a resolved source at these wavelengths -- consistent with the Keck II results of
\citep{ResslerBarsony2001}.
}
\label{fig:radprofiles}
\end{figure*}

In order to determine whether or not any of the continuum sources are extended, we
used the PSF standard observations of 10 Lac (PID 3779;  PI:  D. Gasman) with which to compare
radial profiles of WL 20E, WL 20W, and WL 20SW. 
WL 20SE has too low signal-to-noise and is too confused
with WL 20SW to state one way or the other as to whether or not it is extended.
The azimuthally averaged cross-cuts at 6.2 $\mu$m and 16.6 $\mu$m of
10 Lac, WL 20E, WL 20W, and WL 20SW, each scaled to their individual peak pixel value,
are shown in Figure \ref{fig:radprofiles}. Examination of this figure
demonstrates that whilst WL 20E and WL 20W are consistent with being point sources,
WL 20SW is definitely extended at these wavelengths.

Continuum fluxes for each component of the WL 20 quadruple system are listed in Table \ref{table:results2}.
The MIRI MRS fluxes are measured at the median wavelengths of the SHORT grating in each of the four MIRI MRS channels.
The apertures through which the fluxes are measured have diameters that vary with wavelength as 
2.0 $\times$ FWHM$_{PSF}$ for all sources except for WL 20E, for which the aperture diameter was varied as  3.0 $\times$ FWHM$_{PSF}$,
where FWHM$_{PSF}$ is given by $0.033 (\lambda / \mu m)\  +\  0.106^{\prime\prime}$  \citep{law2023}. 
The larger apertures were necessary for WL 20E in order to minimize fringing at the shortest wavelengths, since WL 20E was very close to the edge of the detector for Channel 1. 
 ALMA Band 6 (1.3 mm) and  Band 4 (1.9 mm) continuum fluxes, and the derived dust disk masses, are also tabulated in Table  \ref{table:results2}; how these were arrived at
 is detailed in $\S$\ref{subsec:ALMA_Band4}.

\begin{deluxetable*}{l   c  c  c  c  c  c   |c c c c c c}
\tabletypesize{\scriptsize}
\tablecolumns{7}
\tablewidth{0pt}
\tablecaption{The WL 20 System:  Coordinates \label{table:results1} }
\tablehead{ 
\colhead{Source} &  \multicolumn{6}{c}{Source Coordinates (MIRI MRS)}   & \multicolumn{6}{c}{Source Coordinates (ALMA Bands 4 \& 6)} \\[-8pt]   
\colhead{Name}   &  \multicolumn{3}{c}{ $\alpha$ (2000) } & \multicolumn{3}{c}{ $\delta$ (2000)}&\multicolumn{3}{c}{ $\alpha$ (2000) }&\multicolumn{3}{c}{ $\delta$ (2000)} \\[-8pt]  
\colhead{}             & \colhead{h} & \colhead{min}  & \colhead{sec}  & \colhead{ $^{\circ}$ }     & \colhead { $^{\prime}$ }  &  \colhead{ $^{\prime\prime}$ } & \colhead{h} & \colhead{min}  & \colhead{sec}  & \colhead{ $^{\circ}$ }     & \colhead { $^{\prime}$ }  &  \colhead{ $^{\prime\prime}$ } 
}
\startdata
   WL 20E              &16  &27  &15.9075   & $-$24  &38 &43.8180  & 16 & 27 & 15.889   & $-$24 &  38 & 43.977   \\
   WL 20W              &16  &27  &15.666    & $-$24  &38 &43.930    & 16 & 27 & 15.652    & $-$24 & 38  & 43.982    \\
   WL 20SW            &16  &27  &15.686    & $-$24  &38 &46.219   & 16  & 27 & 15.674    & $-$24 & 38  & 46.260    \\
   WL 20SE            &16 &27  & 15.728     & $-$24  & 38& 46.141  & 16  & 27 & 15.713    & $-$24 & 38  & 46.133   \\
\enddata
\end{deluxetable*}

\begin{deluxetable*}{l c c c c c c c    }
\tabletypesize{\scriptsize}
\tablecolumns{5}
\tablewidth{0pt}
\tablecaption{The WL 20 System:  Continuum Fluxes and Disk Dust Masses \label{table:results2} }
\tablehead{ 
\colhead{Source}  &\colhead{Flux at}  &  \colhead{Flux at}        &\colhead{Flux at}                   & \colhead{Flux at}          & \colhead{Flux at}  & \colhead{Flux at}    &\colhead{Dust mass}   \\[-10pt] 
\colhead{Name}   &\colhead{5.3 $\mu$m} &\colhead{8.1 $\mu$m}&\colhead{12.5 $\mu$m}    &\colhead{19.3 $\mu$m}    &\colhead{1.3 mm}  &\colhead{1.9 mm} &     \colhead{} \\[-10pt]
\colhead{}              &\colhead{(mJy) }              &\colhead{(mJy) }              &\colhead{(mJy) }              &\colhead{(mJy)}   &\colhead{ (mJy) } &\colhead{ (mJy)} &\colhead{M$_{\oplus}$ }  
}
\startdata
  WL 20E             & 81$\pm$ 0.5             & 27 $\pm$ 1         & 85  $\pm$ 0.5       &95 $\pm$ 2      & 2.1 $\pm$ 0.2   &1.3 $\pm$ 0.1  & 3.3 $\pm$  0.4           \\
  WL 20W             & 34 $\pm$  2             & 42 $\pm$ 1         & 50  $\pm$ 1          &133  $\pm$ 1     & 2.8 $\pm$ 0.3    &1.4 $\pm$ 0.2 &3.6 $\pm$ 0.5           \\
  WL 20SW             & 14 $\pm$ 2             & 89 $\pm$1         & 47.5  $\pm$ 5    &2300 $\pm$ 5      & 36.1 $\pm$ 0.6 &16.2 $\pm$ 0.3 & 42 $\pm$ 2          \\
  WL 20SE            & 3.7$\pm$ 0.5          & $-$                       & $-$                                   & $-$       & 20.1 $\pm$ 0.6  & 9.0 $\pm$ 0.3  & 24 $\pm$ 4       \\
\enddata
\end{deluxetable*}

\subsubsection{On-source Spectra}

\begin{figure*}
\centering
\includegraphics  [width=\linewidth,    height=3.5cm]  {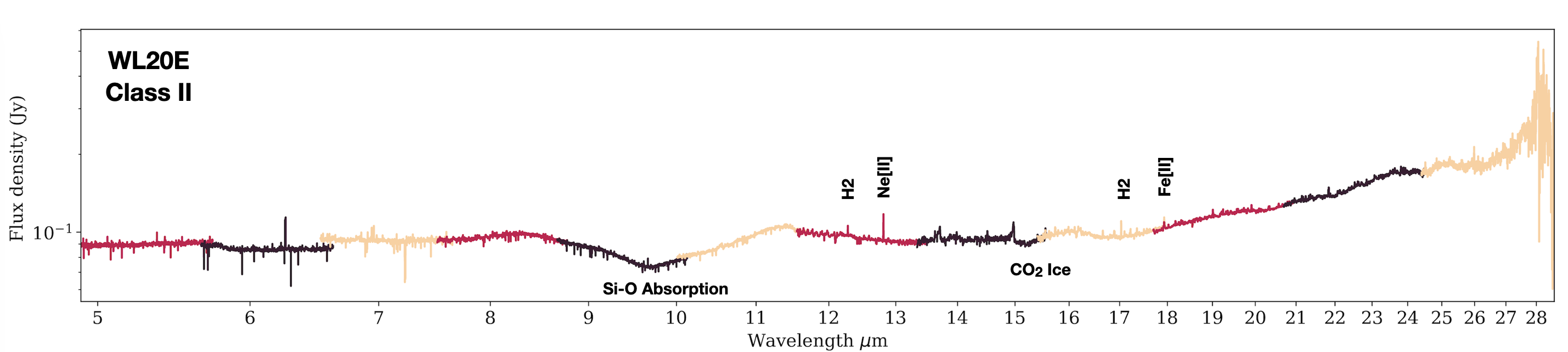}
\includegraphics  [width=\linewidth,  height=3.5cm]  {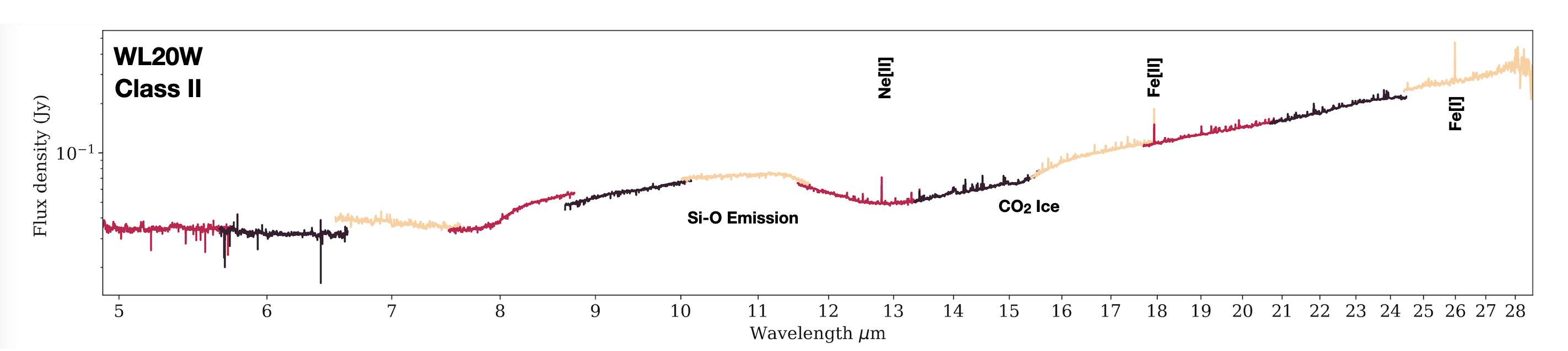}
\includegraphics  [width=\linewidth,   height=3.5cm]  {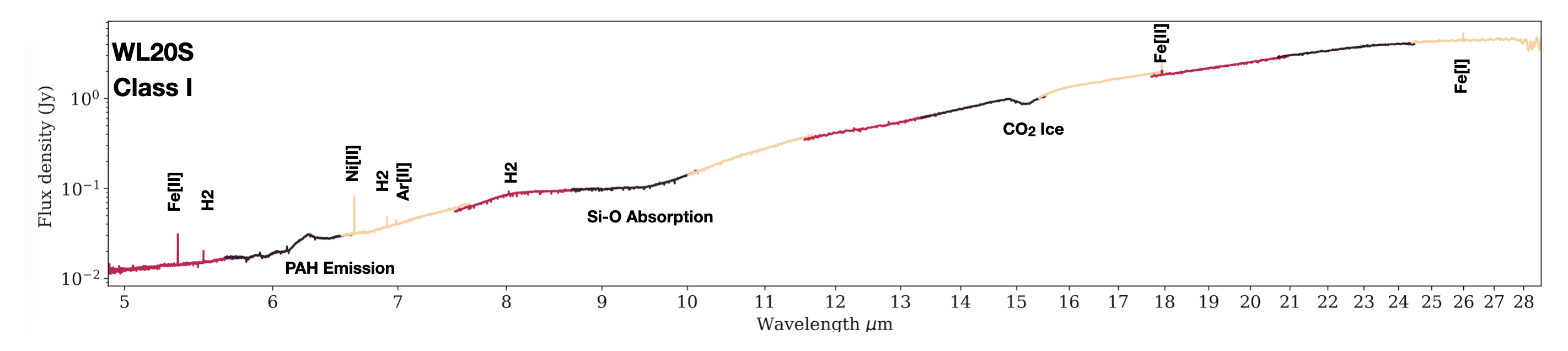}
\caption{5 $\rightarrow$ 28 $\mu$m MIRI MRS spectra of WL 20E (top), WL 20W (middle), and WL 20S (bottom).
The three MRS gratings, A,B, and C are color-coded as red, black, and yellow for all four channels.
Spectra were extracted through apertures scaling as  2.0 $\times$ FWHM$_{PSF}$ for WL 20W and WL 20S,
and as 3.0 $\times$ FWHM$_{PSF}$ for WL 20E. The larger aperture was used for WL 20E to minimize fringing at the shortest wavelengths,
since WL 20E was close to the edge of the detector for Channel 1.
Note the different y-axis scales for the three sources, necessitated by the 
large increase in flux of WL 20S towards longer wavelengths relative to its northern neighbors.}
\label{fig:spectra}
\end{figure*}

On-source MIRI MRS spectra of WL 20E, WL 20W, and WL 20S,
are presented in the top, middle, and bottom panels of Figure \ref{fig:spectra}, respectively. 
Spectra were extracted through apertures scaling as  2.0 $\times$ FWHM$_{PSF}$ for WL 20W and WL 20S,
and as 3.0 $\times$ FWHM$_{PSF}$ for WL 20E. The larger aperture was used for WL 20E in order to minimize fringing at the shortest wavelengths,
since WL 20E was close to the edge of the detector for Channel 1.
Although the aperture through which the WL 20S spectrum was extracted is centered on the coordinates of WL 20SW, 
the spectra of WL 20SW and WL 20SE are blended longwards of 6.0 $\mu$m.

\begin{figure*}
\plottwo{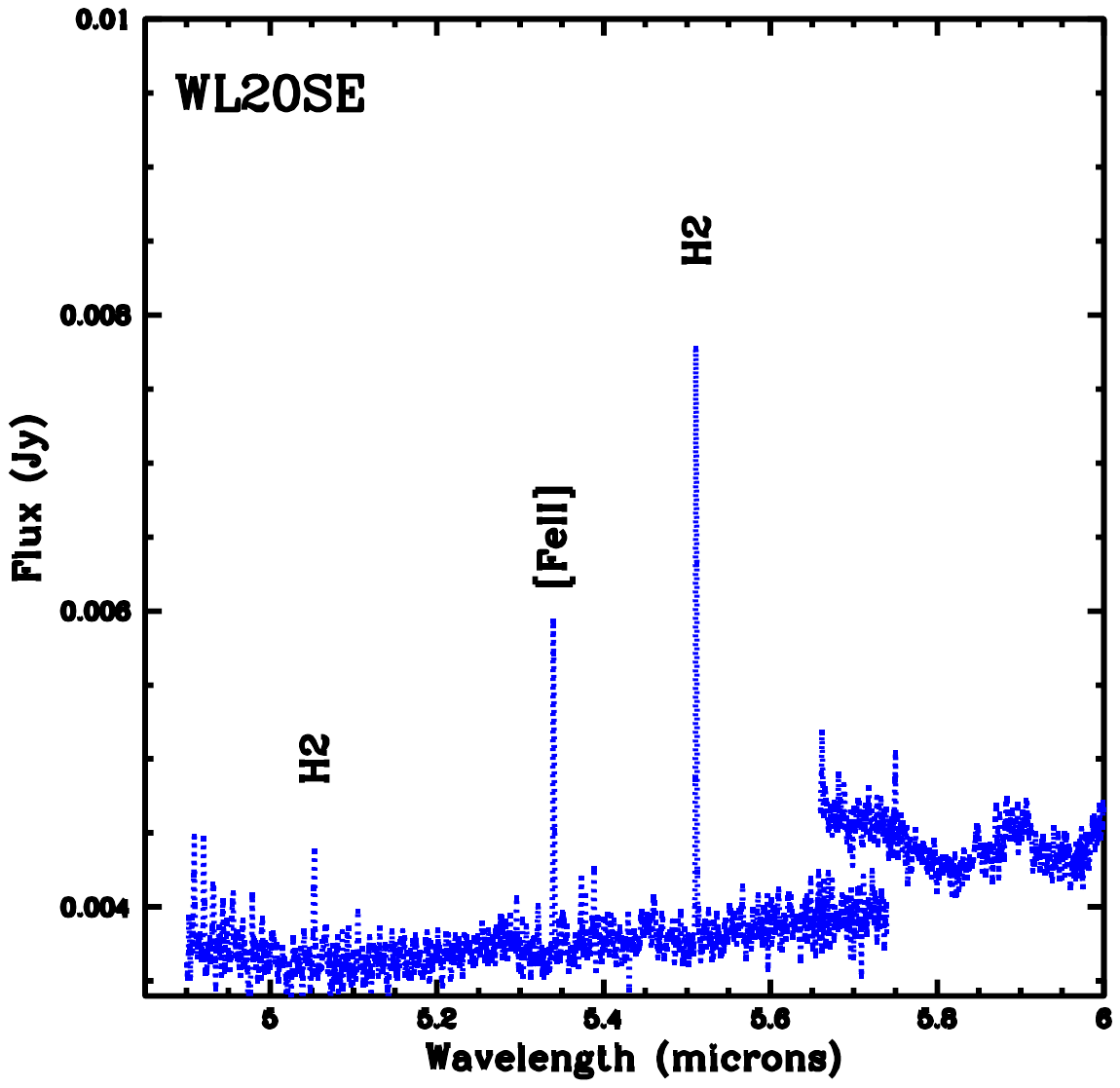}{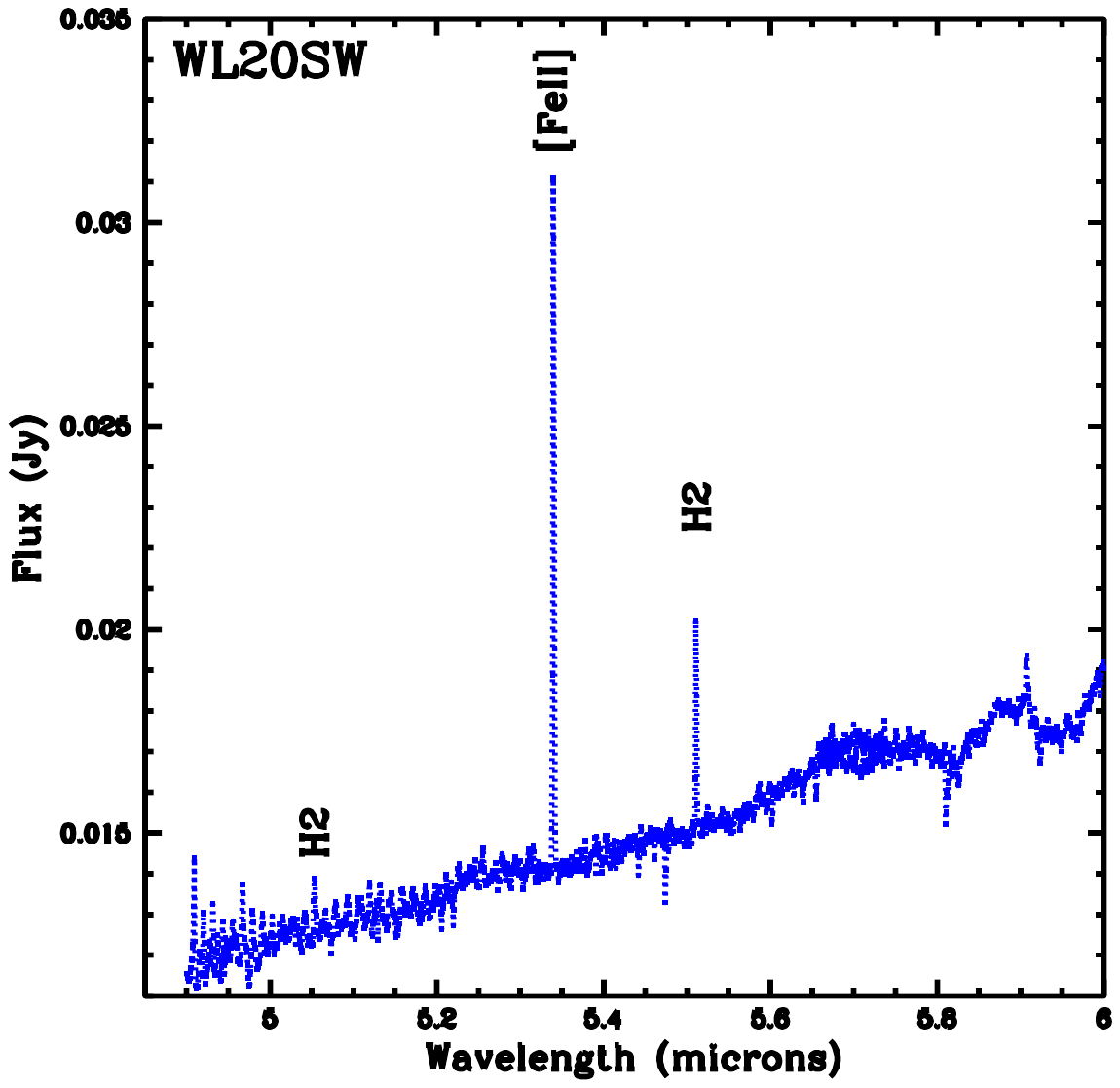}
\caption{  
{\bf Left panel}:  Spatially resolved spectrum of WL 20SE across the 4.9 $\mu$m $-$ 6.0 $\mu$m wavelength
range. A small continuum discontinuity occurs where Channel 1 SHORT and Channel 1 MEDIUM overlap.
{\bf Right panel}:  Spatially resolved spectrum of WL 20SW across the 4.9 $\mu$m $-$ 6.0 $\mu$m wavelength range.
}
\label{fig:resolvedspectra}
\end{figure*}

Spectra of the faint new source, WL 20SE and its bright companion, WL 20SW, are presented in
Figure \ref{fig:resolvedspectra}, in the  wavelength region where the two sources can be spatially resolved.
Note the bright emission lines of H$_2$ and [FeII].
The 5.3 $\mu$m continuum flux level of WL 20SE is a factor
of 3.8 times weaker than that of its neighbor, WL 20SW (see Table \ref{table:results2}).
In the resolved spectra of Figure \ref{fig:resolvedspectra}, WL 20SW displays a steeply rising continuum in contrast
to the relatively flat continuum of its newly discovered neighbor, WL 20SE.

\begin{figure*}
\plottwo{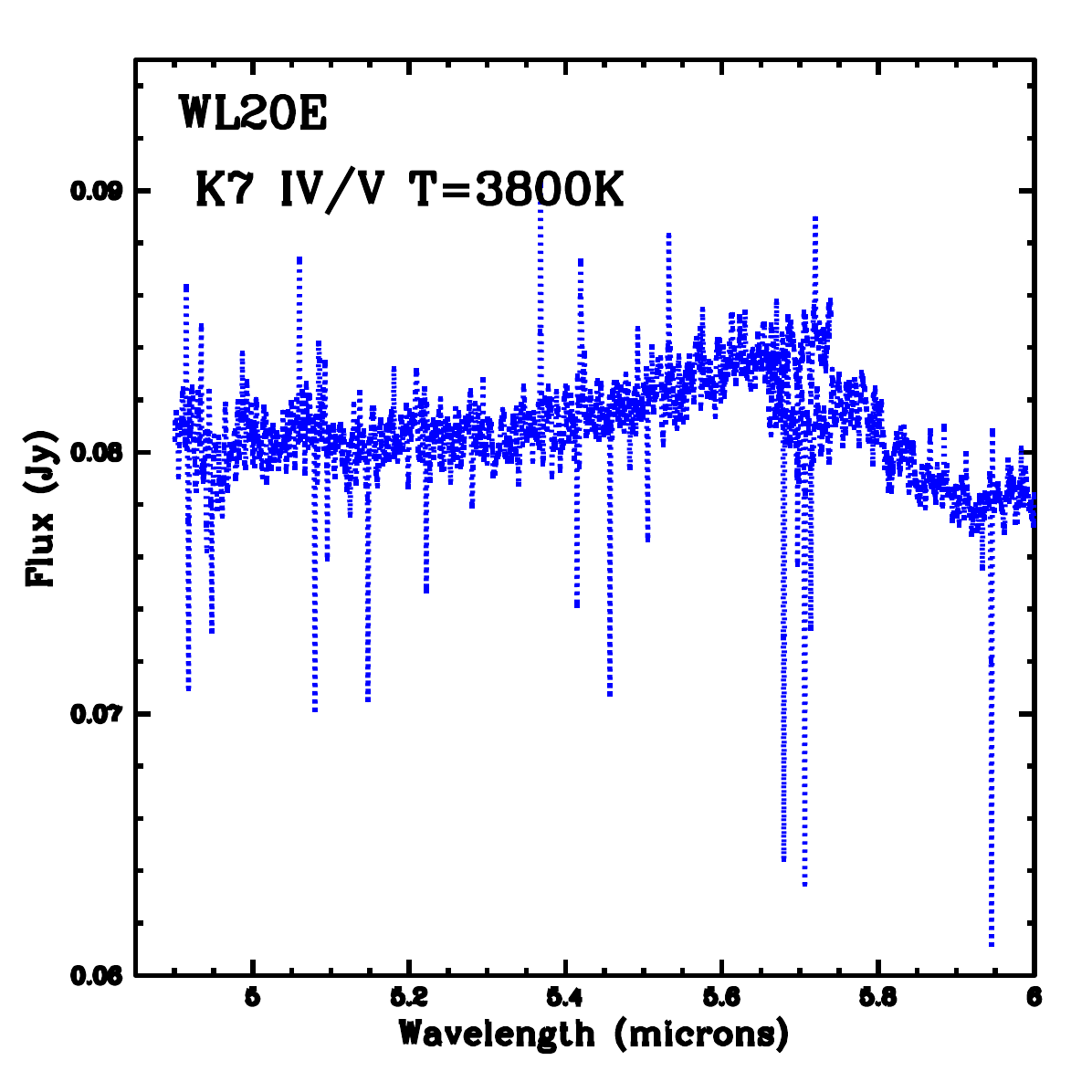}{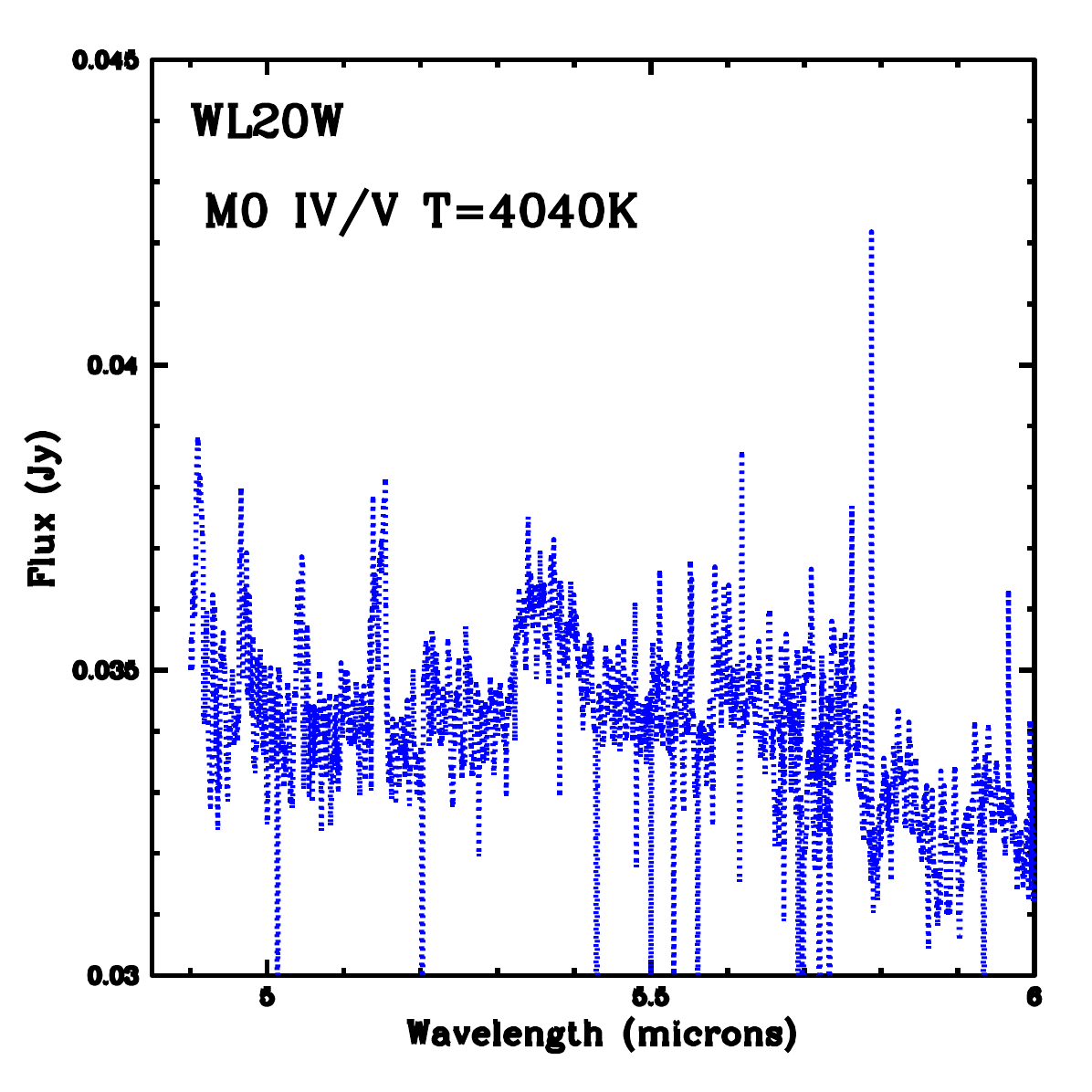}
\caption{  
{\bf Left panel}:  The 4.9 $\mu$m $-$ 6.0 $\mu$m spectrum of WL 20E -- the broad absorption feature in the spectrum of WL 20E
from 5.7 microns till past 6.0 microns is due to H$_2$O ice.
{\bf Right panel}: the 4.9 $\mu$m $-$ 6.0 $\mu$m spectrum of WL 20W -- all of the sharp emission/absorption
features, as well as the broad, emission-like features in the WL 20W spectrum at 5.2 microns and 5.4 microns, are detector artifacts.
Spectral types and temperatures are from \citep{BGB2002}. Note the absence of H$_2$ and [FeII]  emission
in both spectra.
}
\label{fig:WL20eWL20wspectra}
\end{figure*}

Figure \ref{fig:WL20eWL20wspectra} shows the
spectra of WL 20E and WL 20W over this same wavelength interval.  
The spectral types and temperatures shown on their spectra are from \citet{BGB2002}.
Note the much higher continuum fluxes of WL 20E at 81 mJy  and WL 20W 
at 34 mJy in this wavelength region relative to their southern neighbors,
WL 20SW at 14 mJy and WL 20SE at 3.7 mJy, emphasizing the large extinction difference between them.

Also notable is the complete absence of the  [FeII] and H$_2$ emission
lines in the spectra of WL 20E and WL 20W,
in striking contrast to their appearance in the spectra of the WL 20SE/WL 20SW pair.  

In comparison with recent MIRI/MRS spectra of Class 0 and Class I protostars,
what is remarkable in these spectra are the features which are lacking:
namely, the missing broad, deep absorption features associated with various ices, including
H$_2$O, CO, CH$_3$OH,  CH$_4$, NH$_3$, to name just a few \citep{Yang2022, Federman2024, Narang2024, Nisini2024, Rocha2024, Tychoniecetal2024}.
This is yet another indicator of the relatively advanced, Class II
stage, of the WL 20 multiple system. 

Although there are a wealth of spectral features in each component of the WL 20 system,
the focus of this work is on the [NeII] emission from each, and on the
emission lines found in the WL 20S pair.
Identifying and analyzing all of the spectral lines and solid state features detected in these spectra is left for future investigation.

\subsubsection{Jet Images and Spectra  in Low- and High-excitation Ionic Lines}

\begin{figure*}
\begin{center}
\includegraphics[width=0.3\linewidth, height=7cm]{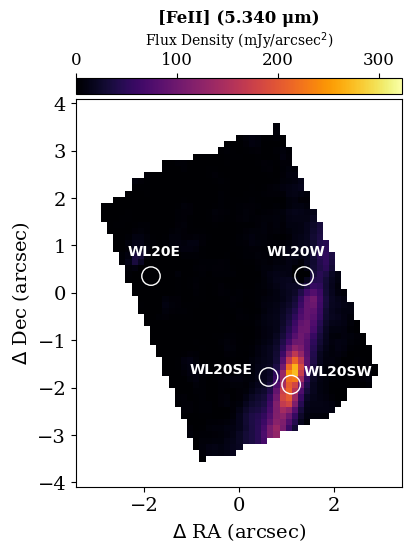}
\quad\includegraphics[width=0.3\linewidth, height=7cm]{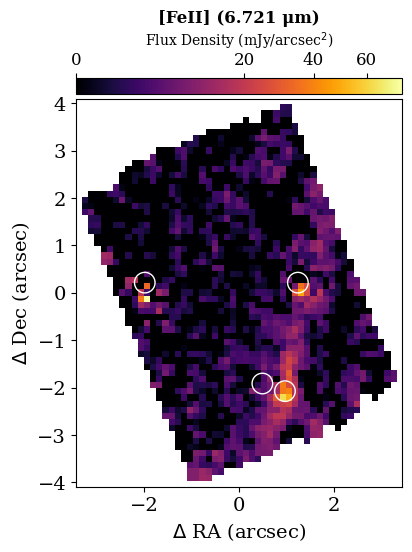}\\[\baselineskip]
\includegraphics[width=.3\linewidth, height=7cm]{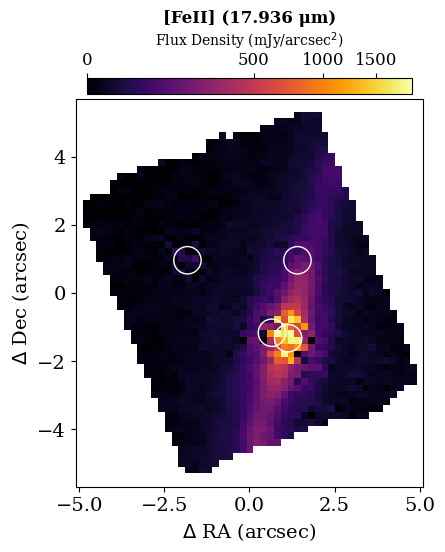}
\quad\includegraphics[width=0.3\linewidth, height=7cm]{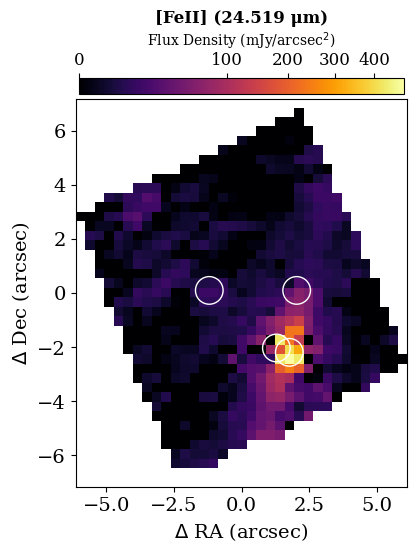}
\quad\includegraphics[width=0.3\linewidth, height=7cm]{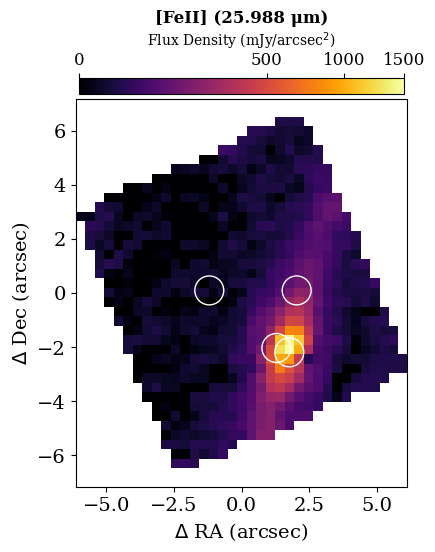}
\end{center}
\caption{Continuum-subtracted, integrated line images of the newly discovered jet emanating from WL 20S in the five transitions of [FeII] listed in Table \ref{table:results4}. Positions of the four continuum mid-infrared components
of the WL 20 system are indicated by open circles, and are labeled in the top-left panel. Color bars indicate the flux scaling in each panel.
Source coordinates are presented in Table \ref{table:results1}.  At an adopted 125 pc distance, 1$^{\prime\prime}$ corresponds to 125 AU at the source.
The FOV of the top panels is 3.2$^{\prime\prime}\,\times\,3.7^{\prime\prime}$. The bottom left panel FOV is 5.5$^{\prime\prime}\, \times\, 6.2^{\prime\prime}$, whereas the 
bottom center and right panels share a 6.9$^{\prime\prime}\, \times\, 7.9^{\prime\prime}$ FOV. 
Artifacts
appearing at the positions of the continuum sources are caused by the undersampling/aliasing of
the spectrograph, which causes pixel-wise spectra to exhibit local flux variations towards continuum
sources, an effect which the line-extraction algorithm has trouble taking into account. 
This effect is particularly visible in the [FeII] 6.721 $\mu$m line image.
}
\label{fig:FeII}
\end{figure*}

\begin{figure*}
\begin{center}
\includegraphics[width=0.3\linewidth, height=7cm]{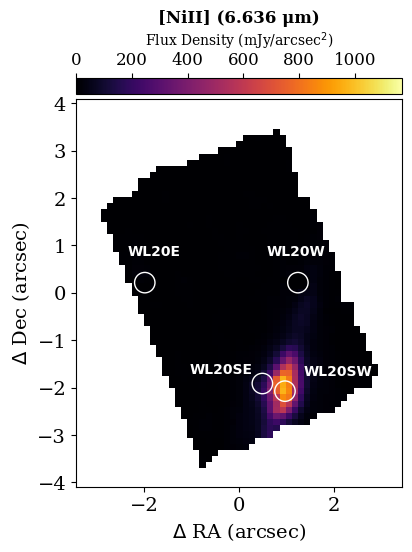}
\quad\includegraphics[width=0.3\linewidth, height=7cm]{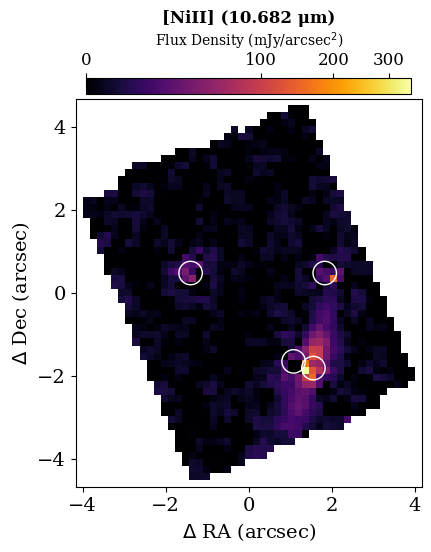}
\caption{Continuum-subtracted, integrated line images of the newly discovered jet emanating from WL 20S in two transitions of [NiII] listed in Table \ref{table:results4}. 
Positions of all four components are indicated by open circles, and labelled in the left panel.  The flux scaling appears in the color bars atop each image.  
One arcsecond corresponds to 125 AU at the source.
The FOV of the left panel is 3.2$^{\prime\prime}\,\times\,3.7^{\prime\prime}$, whereas the FOV of the right panel is 4.0$^{\prime\prime}\,\times\,4.8^{\prime\prime}$.}
\label{fig:NiII}
\end{center}
\end{figure*}

\begin{figure*}
\begin{center}
\includegraphics[width=0.3\linewidth, height=7cm]{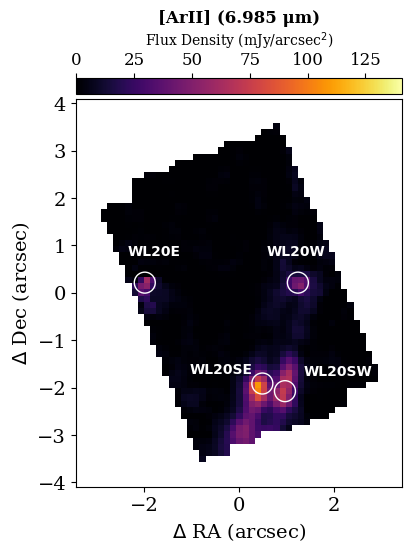}
\quad\includegraphics[width=0.3\linewidth, height=7cm]{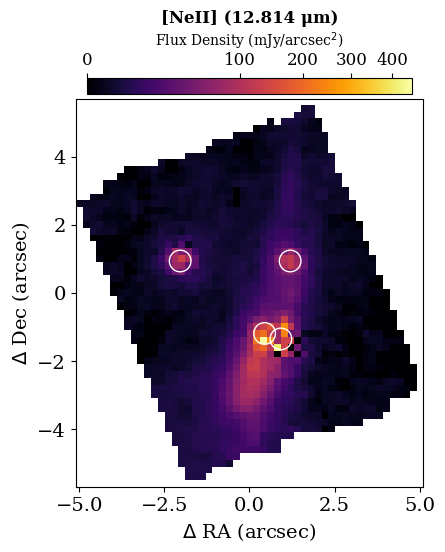}
\label{fig:highexlines}
\caption{Continuum-subtracted, integrated line images of the high-excitation jets seen in the [ArII] and [NeII] lines. 
Positions of all four components of the WL 20 system are indicated by open circles, and are 
labeled in the left panel. 
Flux scales are displayed atop each image by the color bars. At the adopted 125 pc distance to the system, 1$^{\prime\prime}\,=$ 125 AU.
The FOV of the left panel is 3.2$^{\prime\prime}\,\times\,3.7^{\prime\prime}$, whereas the FOV of the right panel is 5.2$^{\prime\prime}\,\times\, 6.2^{\prime\prime}$.
Artifacts
appearing at the positions of the continuum sources are caused by the undersampling/aliasing of
the spectrograph, which causes pixel-wise spectra to exhibit local flux variations towards continuum
sources, an effect which the line-extraction algorithm has trouble taking into account. 
}
\label{fig:hiexlines}
\end{center}
\end{figure*}

A completely unexpected discovery from the MIRI MRS imaging data was that of twin, parallel
jets, powered by the sources WL 20SW and WL 20SE in both low- and high-excitation ionic lines.
Table \ref{table:results4} lists the ionic transitions, excitation potentials, wavelengths, spectral resolutions at these wavelengths, and velocity
extents of the jets.  Continuum-subtracted, emission-line images of these twin jets, each powered by one component of the WL 20S binary, are
presented in Figure \ref{fig:FeII} for the [FeII] lines, in Figure \ref{fig:NiII} for the [NiII] lines,
and Figure \ref{fig:hiexlines} in the [ArII] and [NeII] lines.  

Surprisingly, WL 20SE, the faint, newly discovered source, drives jets that have stronger emission in the
highly excited [ArII] and [NeII] lines than the jets in these lines emanating from its brighter neighbor, WL 20SW.
The situation is reversed in the lower-excitation [FeII] and [NiII] lines, in which the WL 20SE jets 
are fainter than the corresponding jets seen in these lines driven by WL 20SW.

\begin{figure*}
\plotone{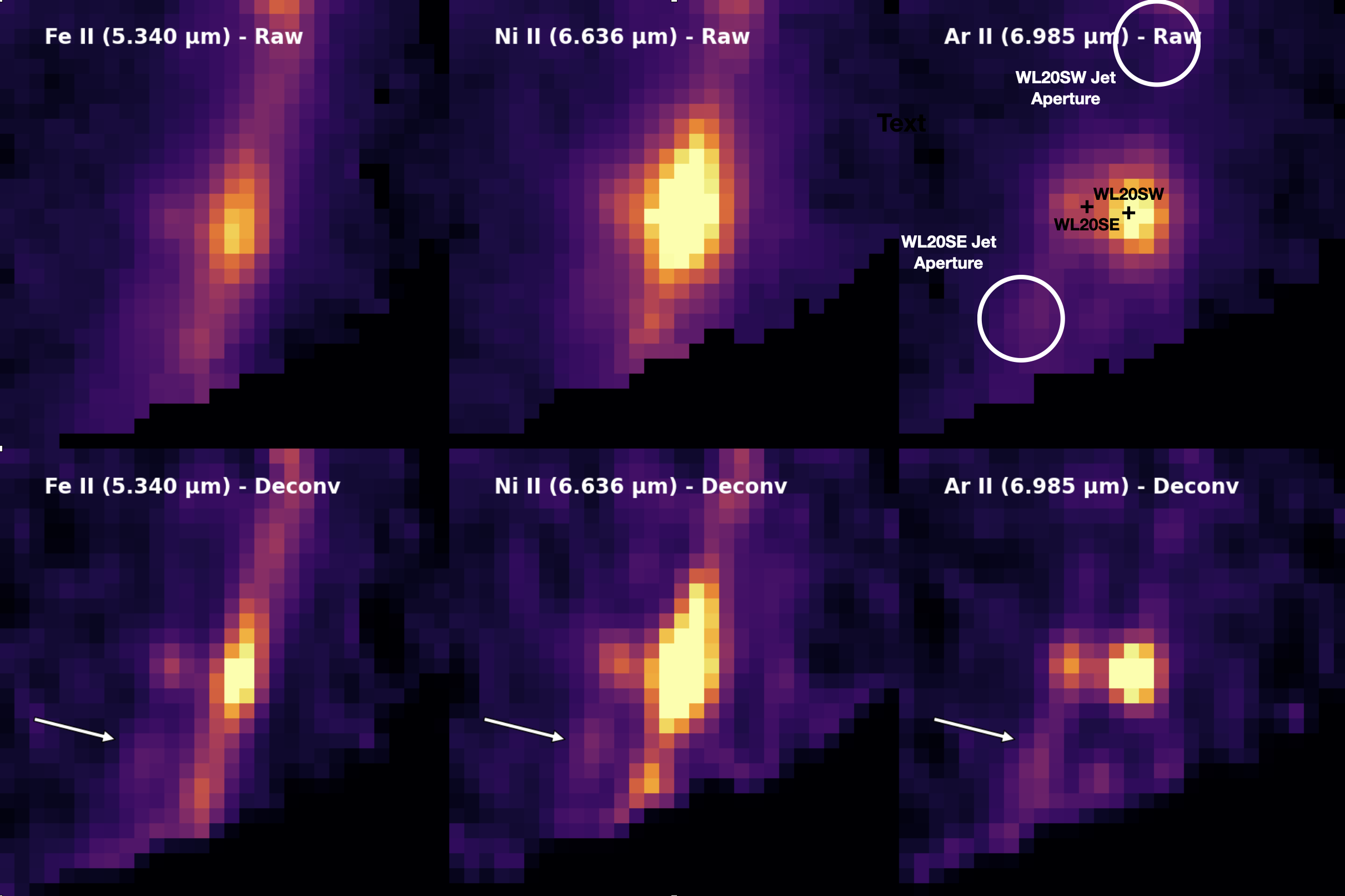}
\caption{Top panels:  Zoomed-in views of the WL 20SE and WL 20SW jets in the continuum-subtracted [FeII] (5.340 $\mu$m),
[Ni II] (6.636 $\mu$m), and [ArII] (6.985 $\mu$m) lines. The 5.3 $\mu$m continuum source positions are marked with the black plus signs in the right panel.
White circles of 1.00$^{\prime\prime}$ diameter mark the apertures used for extracting the spectra in \ref{fig:jetspectra}.
Bottom panels:  Same as in the top panels, but deconvolved, after 16 iterations using the Lucy-Richardson algorithm
as coded in the scikit-image package \citep{vanderwalt2014}.  The deconvolution kernel was derived from wavelength-equivalent observations
of 10 Lac (PID 3779, PI D. Gasman). 
Pixel sizes in all images are 0.196$^{\prime\prime}$.  A square-root intensity stretch with the data normalized from zero to 1 (set to the data maximum) 
was used to bring out the low-level details.
The southern  jet driven by WL 20SE is clearly visible, and is indicated by the white arrows.
} 
\label{fig:deconvolved}
\end{figure*}

Since these results are difficult to see from the jet images of Figures \ref{fig:FeII}, \ref{fig:NiII}, and \ref{fig:hiexlines}, 
in order to best distinguish the separate jets driven by WL 20SE and WL 20SW,
we present zoomed-in views of the twin jets in Figure \ref{fig:deconvolved}.
The top panel shows the unprocessed, continuum-subtracted line images
of [FeII]  (5.340 $\mu$m), [NiII] (6.636 $\mu$m), and  [ArII] (6.985 $\mu$m).
These transitions were chosen because they are at the shortest wavelengths in the  low- and high-excitation jets,  
where the angular resolutions are best. The bottom panel of Figure \ref{fig:deconvolved} shows the 
corresponding deconvolved images, in which the faint southern jet, driven by WL 20SE,
is clearly seen and is indicated by the white arrows.

To further highlight the differences in excitation of the jets, additional jet spectra were obtained. The jet spectra
were extracted from carefully
sized and placed apertures to minimize cross-contamination between the jets.  The locations and sizes of these apertures are depicted by white circles
on the continuum-subtracted 6.985 $\mu$m [Ar II] line image in the top-right panel of Figure \ref{fig:deconvolved}.
The central coordinates of the circular, 1.00$^{\prime\prime}$ diameter apertures used to extract the jet spectra
are listed in Table \ref{table:jetcoords}.
For the WL 20SW jet, the chosen aperture is to the northwest of WL 20SW, to best isolate its jet emission
from the WL 20SE jet's emission. For the WL 20SE jet, the chosen aperture is located to the southeast of WL 20SE, 
as far as
possible from the southern jet driven by WL 20SW, to avoid contamination by WL 20SW's southern jet, but still capturing as 
much as possible of the emission from WL 20SE's jet.
These apertures were 
chosen so that they are separated by the FWHM$_{PSF}$ even at emission lines detected at MIRI's longest wavelengths.
Nevertheless, there will be the inevitable contamination of the continuum levels of the extracted jet spectra, which increases
with wavelength. 
 
\begin{figure*}
\plotone{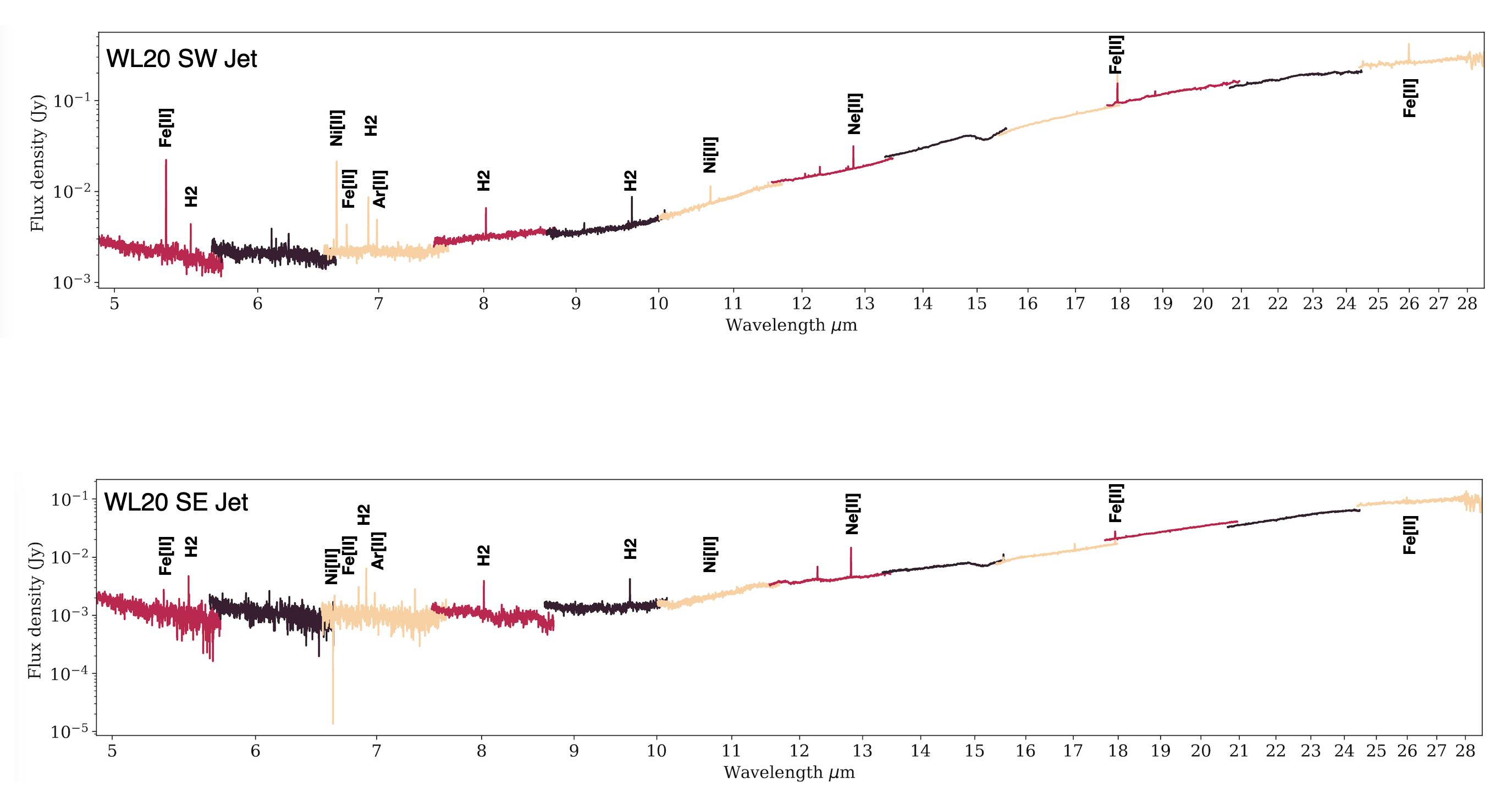}
\caption{Spectra extracted from 1.00$^{\prime\prime}$ diameter circular apertures centered on the coordinates listed in Table \ref{table:jetcoords}
and pictured in Figure \ref{fig:deconvolved} for the WL 20SW jet (top panel) and the WL 20SE jet (bottom panel). Red, black, and yellow correspond to the three MIRI MRS gratings, A,B, and C for all four MIRI MRS channels.
Emission lines in which jets are imaged in Figures \ref{fig:FeII} through \ref{fig:hiexlines} are labelled, as 
are some of the ambient H$_2$ molecular emission lines.}
\label{fig:jetspectra}
\end{figure*}

 The extracted spectra from these two apertures are presented in Figure \ref{fig:jetspectra}. 
  At the shortest wavelengths, where the spatial resolution allows the best separation of the jet apertures from the continuum emission
 of WL 20SE and WL 20SW, the spectra exhibit a suppressed continuum dominated by emission lines from shocked gas.
 As we progress to longer wavelengths, the inevitable contamination of the continuum levels of the jet spectra increases
 in Figure \ref{fig:jetspectra}. 
Examination of the jet spectra of WL 20SE and WL 20SW in Figure \ref{fig:jetspectra} shows the presence of both low- and high-excitation
emission lines in the jets powered by each source -- a result that is difficult to see from the jet images 
of Figures \ref{fig:FeII} $-$ \ref{fig:hiexlines} alone.  The jet spectra also show the much stronger [ArII] and [NeII] line strengths in the jet driven by
WL 20SE compared with those of the WL 20SW jet.

\begin{deluxetable*}{lcccccc}
\tabletypesize{\scriptsize}
\tablecolumns{7}
\tablewidth{0pt}
\tablecaption{Low- and High-excitation Jet Emission Lines from WL 20SW and WL 20SE \label{table:results4} }
\tablehead{ 
 \colhead{Line} & \colhead{Wavelength} & \colhead{Transition} & \colhead{Excitation} & \colhead{Ionization} & \colhead{Spectral} & \colhead{Velocity}\\[-10pt ]  
 \colhead{ }        & \colhead{ ($\mu$m) }  & \colhead{ }               & \colhead{Potential}   & \colhead{Potential}    & \colhead{Resolution} & \colhead{Extent}\\[-10pt]
 \colhead{}         & \colhead{}                   & \colhead{}                 & \colhead{(eV)}         & \colhead{(eV)}             &  \colhead{km s$^{-1}$\tablenotemark{a} } & \colhead{ km s$^{-1}$ } 
 }
\startdata
  $[$FeII$]$   &    5.340169  & a4F9/2$-$a6D9/2  & 7.90 & 16.19  &  $\sim$ 83  & $-76.8\rightarrow +102.8$   \\
 $ [$FeII$]$   &    6.721283  & a4F9/2$-$a6D7/2  & 7.90 &  16.19 &  $\sim$ 83   & $-3.7\rightarrow + 67.7$ \\
 $ [$FeII$]$   &  17.935950  & a4F7/2$-$a4F9/2  & 7.90 & 16.19  &   $\sim$ 150  &$-183.0\rightarrow + 117.9$ \\
 $ [$FeII$]$   &  24.519250  & a4F5/2$-$a4F7/2  & 7.90 & 16.19  &   $\sim$ 180  &$-149.8\rightarrow -3.0$  \\
$ [$FeII$]$   &  25.988290  & a6D7/2$-$a6D9/2 & 7.90 & 16.19   &   $\sim$ 200  & $-130.2\rightarrow +77.4$ \\
$  [$NiII$]$   &     6.636000  & 2D3/2$-$2D5/2     & 7.64 & 18.17  &   $\sim$ 84   & $-90.4\rightarrow +126.5$    \\
 $ [$NiII$]$   &   10.682200  & 4F7/2$-$4F9/2     &  7.64 & 18.27  &  $\sim$ 92  & $-60.3\rightarrow +49.1$    \\
 $ [$ArII$]$   &    6.985274 & 2P1/2$-$2P3/2  & 15.76 & 27.63  &  $\sim$ 84  & $-71.8\rightarrow +65.5$   \\
 $ [$NeII$]$   &    12.813550  & 2P1/2$-$2P3/2 & 21.56 &  40.96 &  $\sim$ 100   & $-53.8\rightarrow +121.7$ \\
\enddata
\tablenotetext{a}{Using $R_{MRS(max)}$ value from Table 3, Column 6 for the central wavelength of each subband from Labiano et al. 2021, except for Channel 4,
where we have used the resolving power listed in the last column of Table 1 in \citet{wells2015}.
}
\tablecomments{Line data are from {\tt https://www.mpe.mpg.de/ir/ISO/linelists}}
\end{deluxetable*}

\begin{deluxetable*}{l c c c c c c}
\tabletypesize{\scriptsize}
\tablecolumns{7}
\tablewidth{0pt}
\tablecaption{Center Coordinates for Extracted Jet Spectra \label{table:jetcoords} }
\tablehead{ 
\colhead{Jet } &  \multicolumn{6}{c}{Jet Aperture Coordinates} \\[-8pt ]  
\colhead{Aperture}   &  \multicolumn{3}{c}{ $\alpha$ (2000) } & \multicolumn{3}{c}{ $\delta$ (2000)} \\[-8pt]  
\colhead{Name}             & \colhead{h} & \colhead{min}  & \colhead{sec}  & \colhead{ $^{\circ}$ }     & \colhead { $^{\prime}$ }  &  \colhead{ $^{\prime\prime}$ }
}
\startdata
   WL 20SW Jet              &16  &27  &15.6686     & $-$24  &38 &45.0694    \\
   WL 20SE Jet            &16 &27  & 15.8045     & $-$24  & 38& 47.0638     \\
\enddata
\end{deluxetable*}

\subsubsection{Molecular Hydrogen Line Images}

\begin{figure*}
\begin{center}
\includegraphics[width=0.2\linewidth, height=5cm]{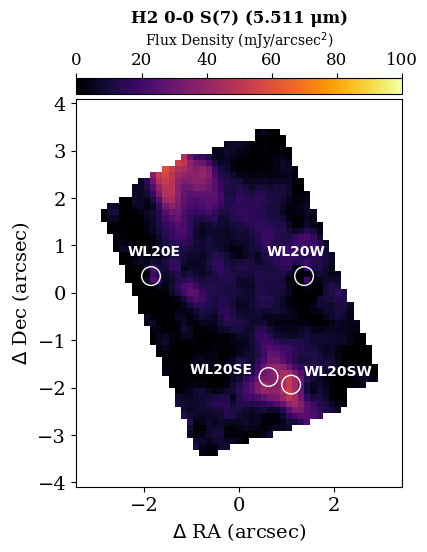}
\quad\includegraphics[width=0.2\linewidth, height=5cm]{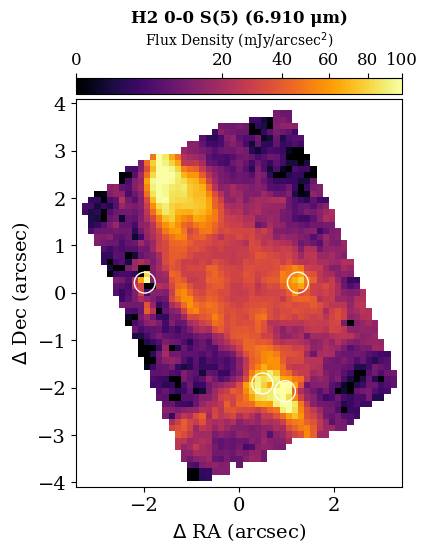}
\quad\includegraphics[width=0.2\linewidth, height=5cm]{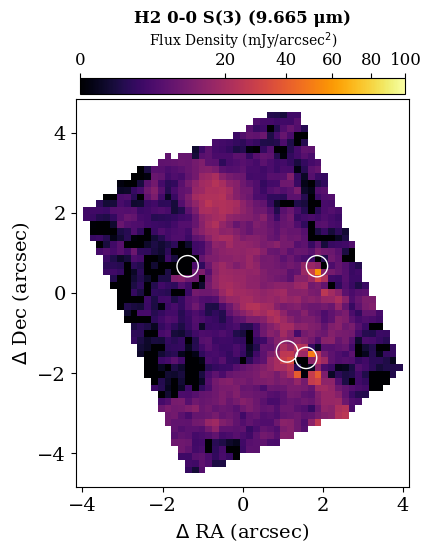}
\quad\includegraphics[width=0.2\linewidth, height=5cm]{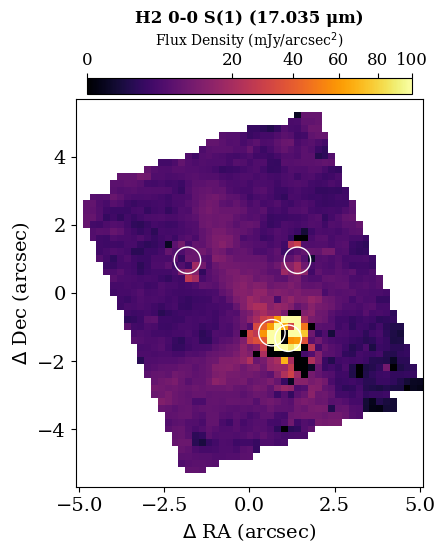}
\\[\baselineskip]
\includegraphics[width=.2\linewidth, height=5cm]{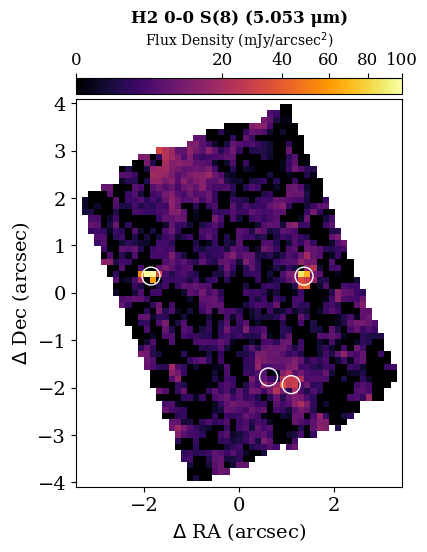}
\quad\includegraphics[width=0.2\linewidth, height=5cm]{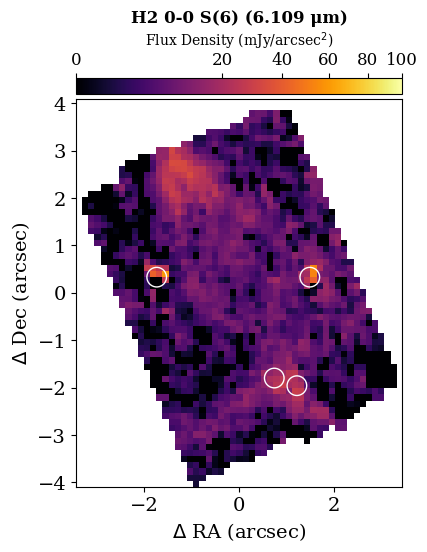}
\quad\includegraphics[width=0.2\linewidth, height=5cm]{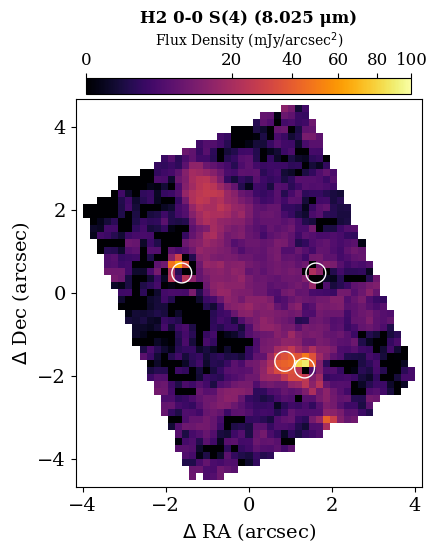}
\quad\includegraphics[width=0.2\linewidth, height=5cm]{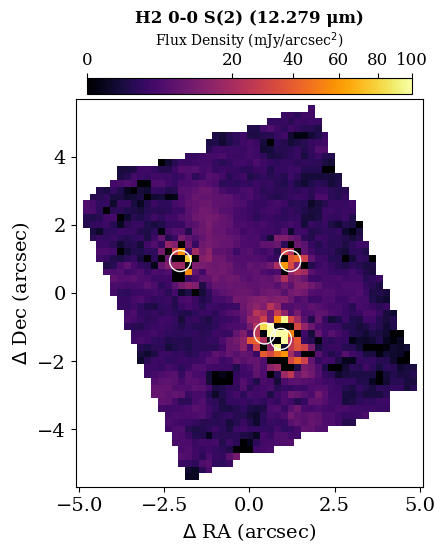}
\end{center}
\caption{The appearance of the WL 20 system in eight distinct molecular hydrogen transitions reveals an X$-$shaped, biconical cone-shaped structure with its apex encompassing both WL 20SE and WL 20SW.
Images in the continuum-subtracted, integrated {\it ortho}-H$_2$ transitions (odd quantum numbers) are in the top row, images in the 
continuum-subtracted, integrated, {\it para}-transitions (even quantum numbers) are in the bottom row.
Open circles mark the locations of the four young stellar objects, labelled in the top-right panel. Color bars indicate the flux scales atop each figure.
At the 125 pc distance to the source, the physical scale is 125 AU to 1$^{\prime\prime}$.
FiOV's vary depending on which of the MIRI MRS channels
each line is detected in: 3.2$^{\prime\prime} \times 3.7^{\prime\prime}$ for 4.90 $\rightarrow$ 7.65 $\mu$m (Channel 1);  4.0$^{\prime\prime} \times 4.8^{\prime\prime}$ for
7.51 $\rightarrow$ 11.7$\mu$m (Channel 2), or 5.2$^{\prime\prime} \times 6.2^{\prime\prime}$ for 11.55 $\rightarrow$ 17.98 $\mu$m (Channel 3).
Artifacts
appearing at the positions of the continuum sources are caused by the undersampling/aliasing of
the spectrograph, which causes pixel-wise spectra to exhibit local flux variations toward continuum
sources, an effect which the line-extraction algorithm has trouble taking into account. 
}
\label{fig:H2maps}
\end{figure*}

Figure \ref{fig:H2maps} displays the distribution of molecular hydrogen line emission in the eight transitions listed in Table \ref{table:H2lines}, continuum-subtracted and 
integrated over all channels in which each line is detected.
Note the stark contrast between the appearance of the gas traced via the molecular hydrogen transitions and the jet-like structures evident in the ionized lines
of Figures \ref{fig:FeII} through \ref{fig:hiexlines}.

\begin{deluxetable*}{rcccccccc}
\tabletypesize{\scriptsize}
\tablecolumns{9}
\tablewidth{0pt}
\centering
\tablecaption{Molecular Hydrogen Lines Detected in the WL 20 System\label{table:H2lines} }
\tablehead{
\colhead{Wavelength} & 
\colhead{Transition} & 
\colhead{$J$} & 
\colhead{$J^{\prime}$}& 
\colhead{ $g_J$ }&
 \colhead{$E_u$}&
 \colhead{$A$}&
 \colhead{Spectral Resolution\tablenotemark{a}       }& 
 \colhead{Velocity Extent}\\[-10pt]
 \colhead{ ($\mu$m) }   &\colhead{}                & \colhead{}     & \colhead{}    &\colhead{}                                & \colhead{(K)}    & \colhead{s$^{-1}$} & \colhead{(km s$^{-1}$)} & \colhead{ (km s$^{-1}$) }
}
\startdata
17.0348   & 0-0 S(1) & 3 & 1 & 21  &  1015& 4.761 $\times 10^{-10}$     & $\sim$ 121  &  $-$63.2  $\rightarrow\ +$24.8\\
 12.2785    & 0-0 S(2) & 4 & 2 &   9 &  1682&  2.755 $\times 10^{-9}$       & $\sim$ 100   & $-$57.6 $\rightarrow\ +$3.4  \\
 9.66491    & 0-0 S(3) & 5 & 3 & 33 &  2504&  9.836 $\times 10^{-9}$       & $\sim$ 91    &  $-$5.3 $\rightarrow\ +$75.4 \\
  8.02505   & 0-0 S(4) & 6 & 4 & 13 &  3475&  2.643 $\times  10^{-8}$      & $\sim$ 85   &   $-$82.2 $\rightarrow\ +$15.0 \\
  6.90952   & 0-0 S(5) & 7 & 5 & 45 &  4586&  5.879 $\times  10^{-8}$      &  $\sim$ 84   & $-$83.3 $\rightarrow\ +$55.5  \\
  6.10856   & 0-0 S(6) & 8 & 6 & 17 &  5830&  1.142 $\times  10^{-7}$      &  $\sim$ 83   & $-$86.9 $\rightarrow\ +$30.9  \\
  5.51116    &0-0 S(7) & 9 & 7  & 57 & 7197&   2.001 $\times  10^{-7}$      & $\sim$ 83   & $-$64.2 $\rightarrow\ +$66.4\\
  5.05311   &0-0 S(8) &10& 8  &  21 & 8677&   3.236 $\times  10^{-7}$      & $\sim$ 83  &  $-$42.7 $\rightarrow\ +$4.8 \\
\enddata
\tablenotetext{a}{Using $R_{MRS(max)}$ value from Table 3, Column 6 for the central wavelength of each subband from Labiano et al. 2021}
\end{deluxetable*}

\begin{figure*}
\begin{center}
\includegraphics[width=0.5\linewidth, height=7.5cm]{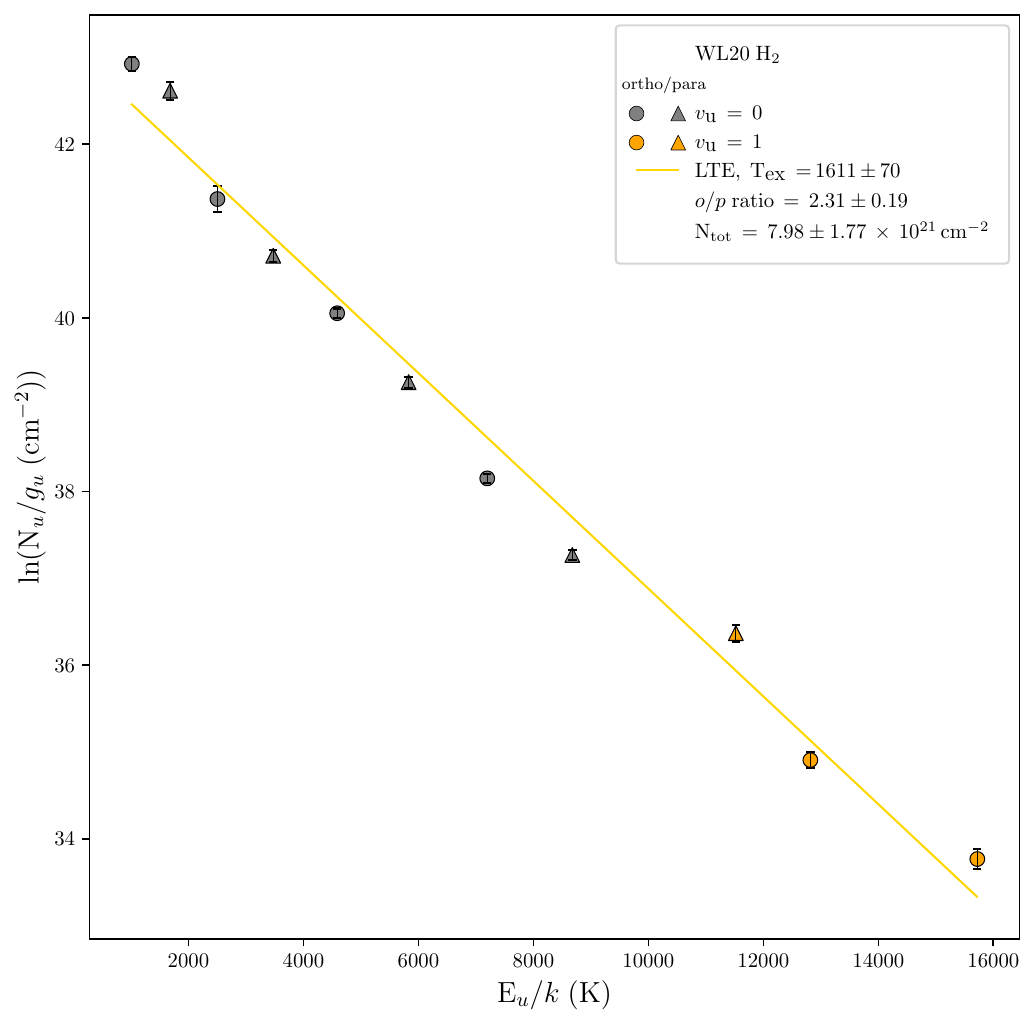}
\end{center}
\caption{Excitation diagram of the H$_2$ emission from the apex of the brightest
H$_2$ S(5) emission to the north of WL 20E, at position $\alpha_{2000}$ 16h 27m 15.85s, 
$\delta_{2000}$ -24$^{\circ}$ 38$^{\prime}$ 41.89$^{\prime\prime}$.
The best-fit line yields the exhibited values of T$_{ex}\, =$ 1611 $\pm$ 70K
and A$_V\, =$ 12 $\pm$ 1.}
\label{fig:excitation}
\end{figure*}

In order to examine the physical conditions of the molecular hydrogen gas, we have produced an excitation
diagram obtained at the apex of the brightest H$_2$ S(5) emission north of WL 20E,
at J(2000) 16h 27m 15.85s, -24$^{\circ}$ 38$^{\prime}$ 41.89$^{\prime\prime}$, simultaneously fitting for 
A$_V$ and T$_{ex}$.  The results are displayed in Figure \ref{fig:excitation}. The derived, best-fit
parameters at this position yield T$_{ex}\,=$ 1161K  $\pm$ 70K, N$_{H2}\,=$ 7.98 $\pm$ 1.77 $\times$ 10$^{21}$ cm$^{-2}$,
and A$_V\,=$ 12 $\pm$ 1.

\subsection{\it ALMA}

\subsubsection{ALMA Band 4 Data}\label{subsec:ALMA_Band4}

\begin{figure*}
\plotone{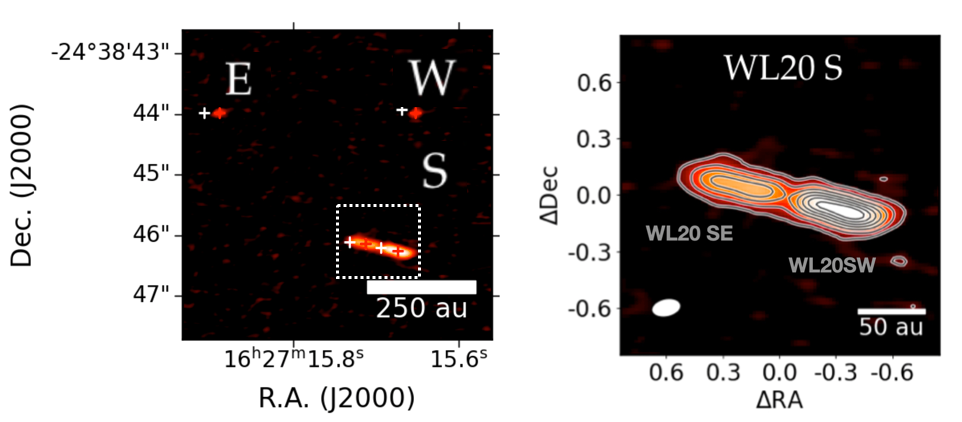}
\caption{Band 4 (1.9 mm) ALMA continuum image of the WL 20 system. 
Source positions listed in Table \ref{table:results1}
are indicated by white (MIRI/MRS) and red (ALMA) crosses, highlighting the small offset between the two.
The dotted rectangle surrounding WL 20S in the left panel is magnified
in the right panel,  highlighting the structure of the newly discovered WL 20S binary.  The source labelled WL 20SE coincides with the position of the newly
discovered mid-infrared source of Figure \ref{fig:WL20continuum}.  Contour levels are at  [3, 10, 20, 30, 40, 50, 60] $\times\ \sigma$ with $\sigma$ = 0.073 mJy/beam.} 
\label{fig:ALMA_Band4}
\end{figure*}

Figure \ref{fig:ALMA_Band4} shows Band 4 (1.9 mm) images of the WL 20 system obtained with ALMA. 
The WL 20S source is resolved into a binary, with the eastern component spatially coinciding with the newly discovered companion
evident at the shortest MIRI MRS wavelengths seen in Figure \ref{fig:WL20continuum}.

The millimeter source positions and continuum fluxes from 
ALMA Band 6 (1.3 mm) and  Band 4 (1.9 mm) are obtained by using the \texttt{imfit} procedure in CASA \citep{McMullin.Waters.ea2007}. 
For WL 20E and WL 20W, each of which remains unresolved by ALMA, a single Gaussian fit was
used for both flux and positional determinations.  By contrast, WL 20SE and WL 20SW are both resolved by ALMA (see Figure \ref{fig:ALMA_Band4}),
and their elongated morphologies are best fit by a combination of two Gaussian components.
The tabulated positions for WL 20SE and WL 20SW are simply the mean value of the two Gaussian fits, whereas the reported fluxes are the sum of the
two Gaussian components fitted to each source. The resulting positions and fluxes are reported in Tables \ref{table:results1} and \ref{table:results2}, respectively.

The continuum fluxes can be used to infer disk {\it dust} masses, assuming optically thin dust emission, using the equation from \cite{Hildebrand1983}:
\begin{equation}
\label{eq:dustmass}
M=\frac{D^2F_\nu}{{\kappa}_\nu B_\nu(T_{\rm dust})}\,
\end{equation}
with $D$ - the distance to the source, $B_\nu$ - the Planck function for a temperature $T_{\rm dust}$, $\kappa_\nu$ - the dust opacity. The temperature of the dust is assumed to be
30 K, typical for young protostellar envelopes \citep{Whitney.Wood.ea2003}. A value of $\kappa_{1.9 {\rm mm}}= 0.6 \ {\rm cm^{2}\ g^{-1}}$ was used, scaled from $\kappa_{1.3 {\rm mm}}= 0.9 \ {\rm cm^{2}\ g^{-1}}$ provided in \cite{Ossenkopf.Henning1994} using $\beta$ = 1. \citep{Andrews.Wilner.ea2009}.  The resulting dust masses are reported
in the last column of Table \ref{table:results2} in units of Earth masses.  

\subsubsection{ALMA Band 6 Data}\label{subsec:Band6}

\begin{figure*}
\begin{center}
\includegraphics[width=0.3\linewidth, height=4cm] {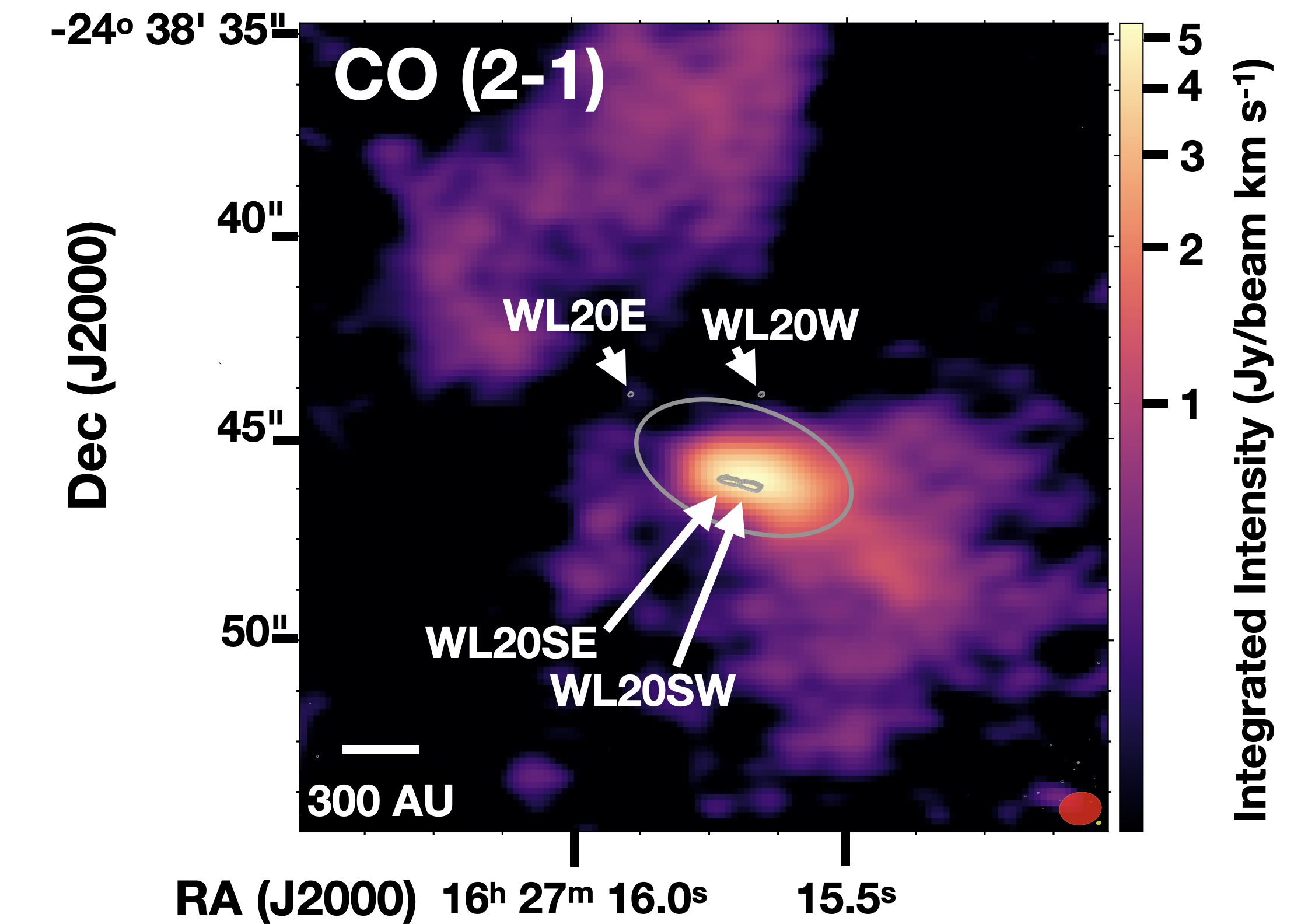}
\quad\includegraphics[width=0.3\linewidth, height=4cm]{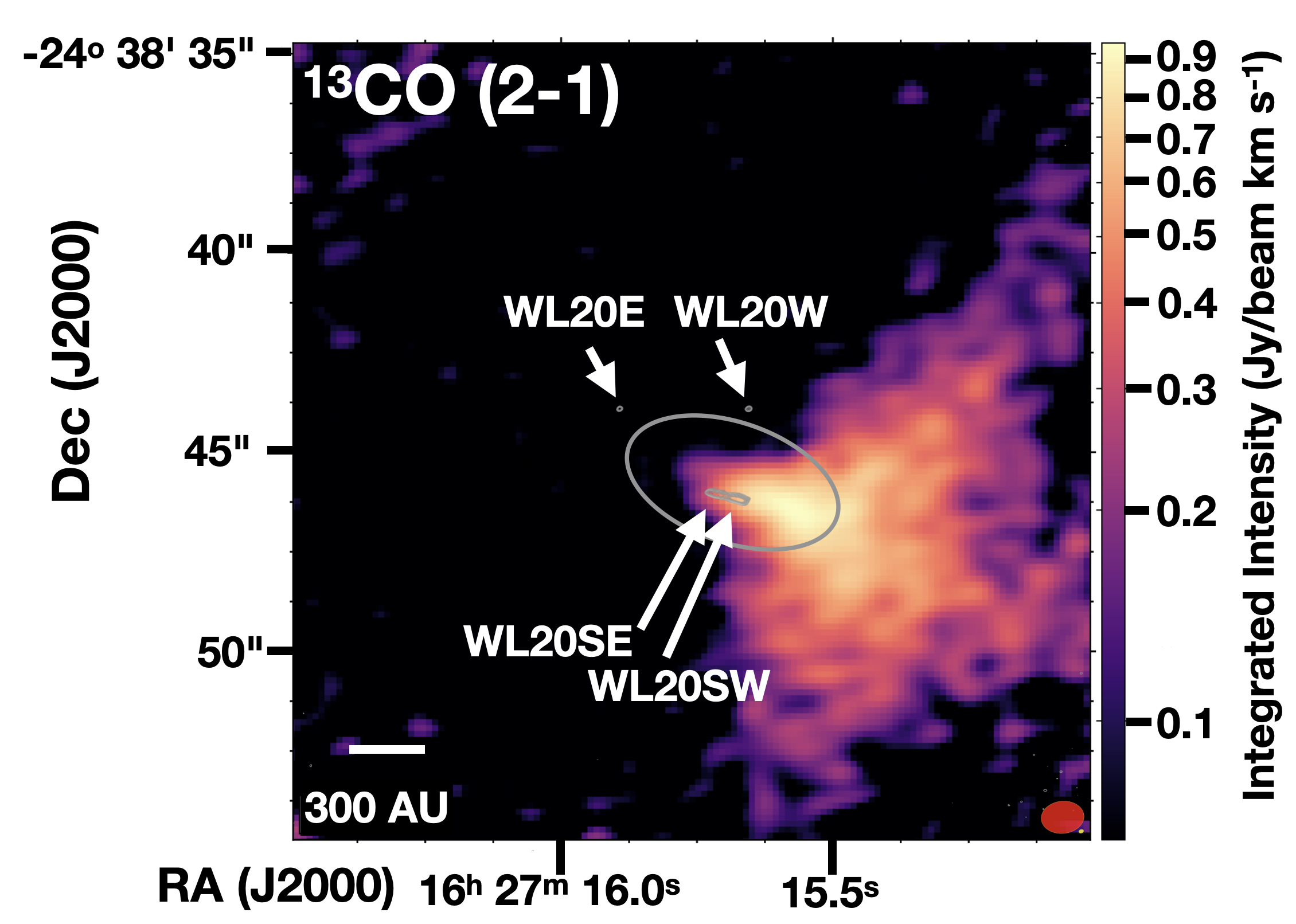}
\quad\includegraphics[width=0.3\linewidth, height=4cm]{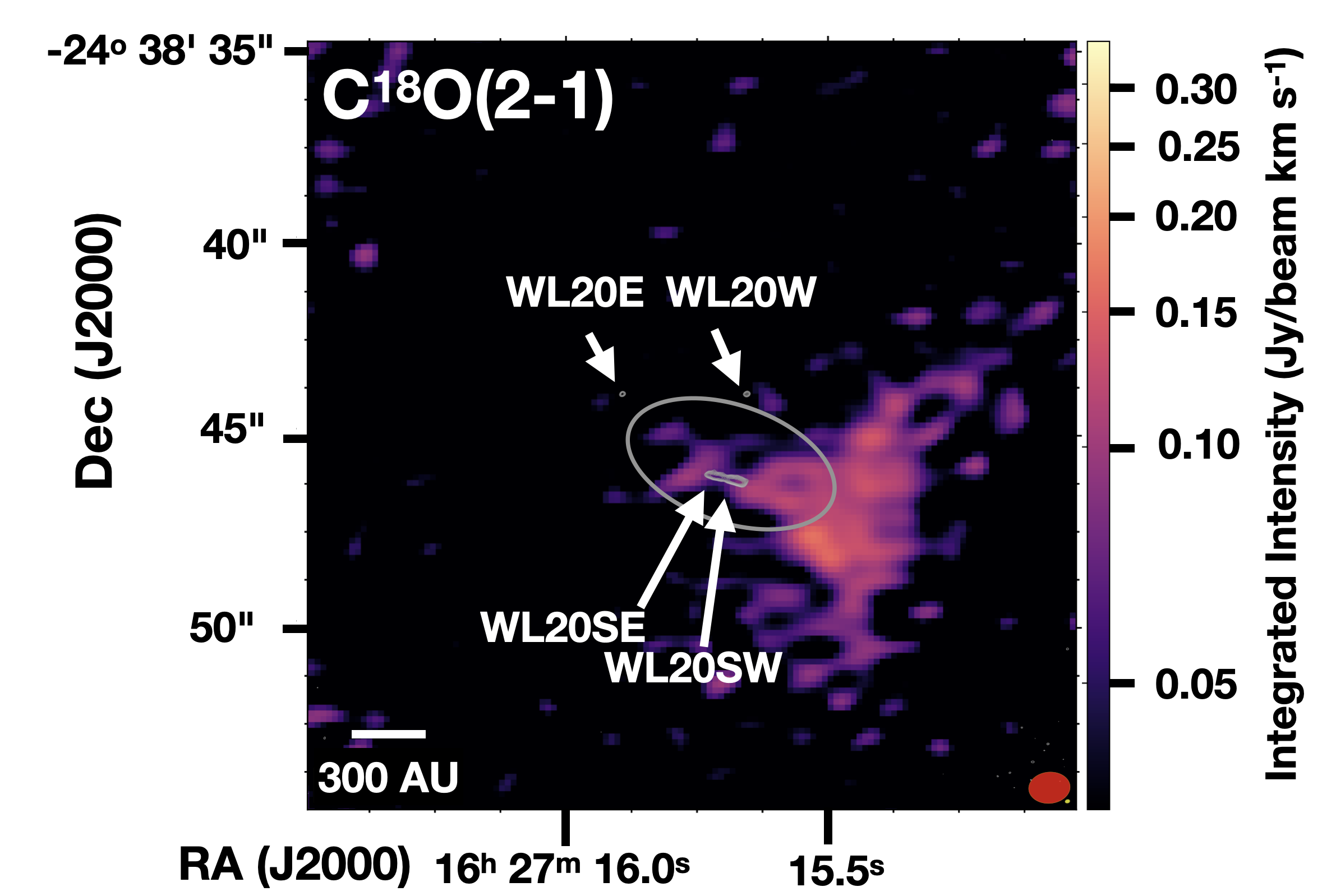}
\end{center}
\caption{ 
ALMA Band 6 observations of the entire WL 20 system.  
The beam FWHM (1.05$^{\prime\prime} \times 0.8^{\prime\prime}$, P.A. $-$83$^{\circ}$)
for the CO isotopologue images is indicated at the bottom right of each panel by the small red ellipse; scale bars are shown at the bottom left of each panel.
Contours represent the  ALMA Band 6 (211 $-$ 275 GHz) continuum emission, acquired at much higher resolution
(FWHM 0.14$^{\prime\prime} \times 0.11^{\prime\prime}$, P.A. $-$75$^{\circ}$). The gray ellipse encompasses the peak CO emission structure centered on the
twin WL 20SE/WL 20SW disks.
{\bf Left panel:} the WL 20 system as imaged in the integrated $^{12}$CO J$=2\rightarrow1$ transition over the $-$2.5 km s$^{-1}$ to $+$12 km s$^{-1}$ velocity range.
{\bf Center panel:} the WL 20 system seen in the $^{13}$CO $J=2\rightarrow1$ transition, integrated over the velocity range $+$1.5 km s$^{-1}\, < v_{LSR}\, <  +9.0$ km s$^{-1}$.
{\bf Right panel:}  the WL 20 system as seen in the C$^{18}$O $J=2\rightarrow1$ transition, integrated over the velocity range 
$+$3 km s$^{-1}\ <\, v_{LSR}=\, +4$ km s$^{-1}\, <\, +$7 km s$^{-1}$.
}
\label{fig:MOM0}
\end{figure*}

\begin{figure*}
\begin{center}
\includegraphics[width=0.3\linewidth, height=4cm] {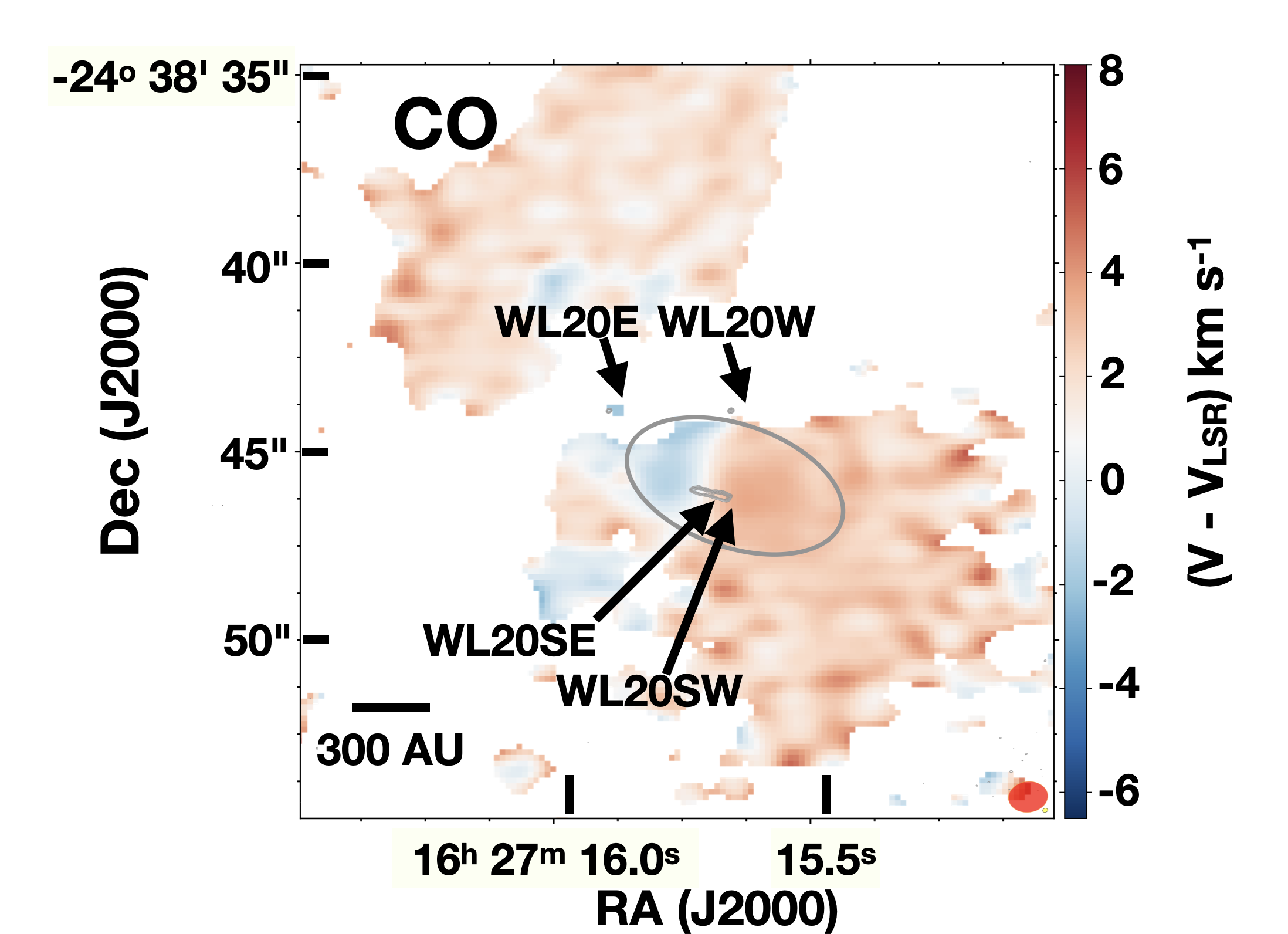}
\quad\includegraphics[width=0.3\linewidth, height=4cm]{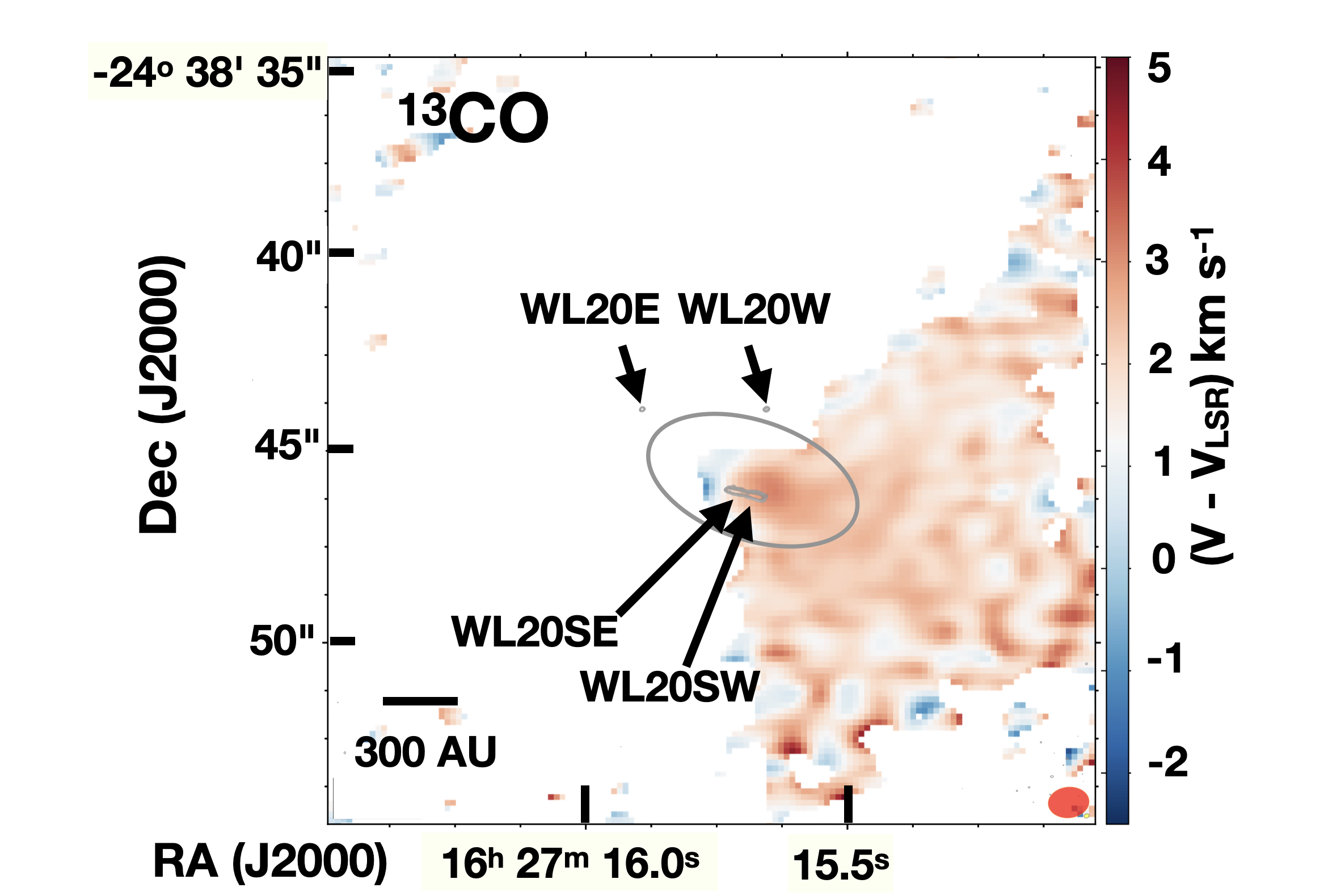}
\quad\includegraphics[width=0.3\linewidth, height=4cm]{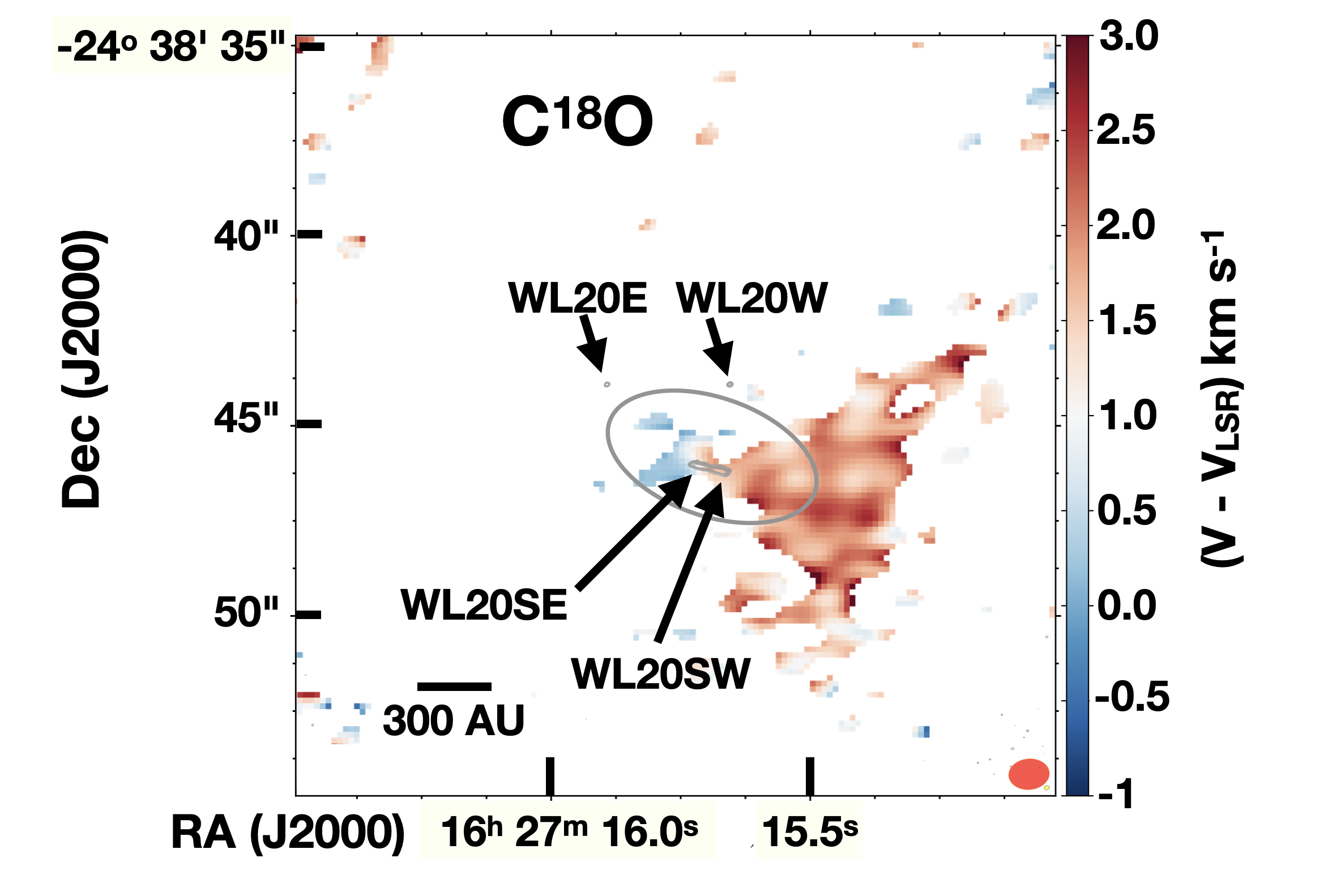}
\end{center}
\caption{ALMA Moment 1 maps in the 2$\rightarrow$1 transitions of CO, $^{13}$CO, and C$^{18}$O of the WL 20 region,
acquired with a beam FWHM of 1.05$^{\prime\prime} \times 0.8^{\prime\prime}$, P.A. $-$83$^{\circ}$,
represented by the red ellipse in the bottom right corner of each panel.
Contours represent the  ALMA Band 6 (211 $-$ 275 GHz) continuum emission of each component of the quadruple system,
acquired with a beam FWHM  of 0.14$^{\prime\prime} \times 0.11^{\prime\prime}$, at P.A. $-$75$^{\circ}$. Scale bars are included in each image.
The gray ellipse in each panel represents the location of the peak CO(2$\rightarrow$1) emission evident in Figure \ref{fig:MOM0}, and exhibits signatures of 
rotational motion in the CO(2$\rightarrow$1) MOM1 map at left (see text for discussion).}
\label{fig:MOM1}
\end{figure*}

Band 6 (1.3 mm) continuum ALMA data were acquired at similar resolution to the Band 4 (1.9 mm) data, and are 
represented by gray contours appearing in Figure \ref{fig:MOM0}.
Positions and fluxes of the sources detected in the  Band 6 continuum are presented in Tables \ref{table:results1} and \ref{table:results2}, respectively.

Figure \ref{fig:MOM0} shows the appearance of the WL 20 system in $^{12}$CO $J=2\rightarrow1$ (left panel),
$^{13}$CO $J=2\rightarrow1$ (middle panel), and C$^{18}$O $J=2\rightarrow1$ (right panel), all shown in color,
with their respective flux scales indicated by the color bars  in Jy/beam km s$^{-1}$ units. 
The FWHM beam size used for the CO isotopologue observations was roughly a factor of 10 larger in each dimension
than the beam size used for the continuum observations, as indicated in the Figure \ref{fig:MOM0} caption.
The $^{12}$CO $J=2\rightarrow1$ emission appears to peak near the elongated continuum structures associated with WL 20SW and its newly discovered companion, WL 20SE.
The gray ellipse drawn in each panel of Figure \ref{fig:MOM0} indicates the location and extent of this CO $J=2\rightarrow1$  peak,
for direct comparison with the gas morphologies  evident in the $^{13}$CO $J=2\rightarrow1$
and  C$^{18}$O $J=2\rightarrow1$ maps, as well as with the Moment 1 maps presented in Figure \ref{fig:MOM1}.

Although the ALMA CO observations do not have enough angular resolution to resolve
the twin disks of WL 20SE and WL 20SW individually, we can, nevertheless, determine velocity-integrated 3$\sigma$ line fluxes from the $^{13}$CO(2$-$1) and C$^{18}$O(2$-$1)
moment-zero maps for later use in determining the total gas mass associated with both disks together.
When measured within two beam areas (0.46$^{\prime\prime}\ \times$ 0.92$^{\prime\prime}$ at P.A. $=$ 345$^{\circ}$),
centered on the two edge-on disks detected in the continuum,
the $^{13}$CO(2$-$1) and C$^{18}$O(2$-$1) line fluxes
are 0.6869 Jy km s$^{-1}$ and 0.09979 Jy km s$^{-1}$, respectively.

In contrast to the CO $J=2\rightarrow1$ peak associated with the twin disks of WL 20SW $+$ WL 20SE,
there is a remarkable lack of gas emission toward either of the sources, WL 20E or WL 20W.  Quantitative upper limits
on this emission are derived by measuring the 3 $\sigma$ rms in a line-free region of the $^{13}$CO(2$-$1) moment-zero
map, which yields a value of $\le$ 1.11 $\times$ 10$^{-3}$ Jy km s$^{-1}$,  and a corresponding 3 $\sigma$ upper limit for the C$^{18}$O(2$-$1) emission $\le$ 5.1 $\times$ 10$^{-4}$ Jy km s$^{-1}$.

\section{Discussion}\label{sec:discuss}

\subsection{\it MIRI MRS Continuum Sources}\label{subsec:MIRIdiscuss}

\begin{deluxetable*}{lccccccc}
\tabletypesize{\scriptsize}
\tablecolumns{8}
\tablewidth{0pt}
\tablecaption{  Continuum Flux Comparison of the WL 20 Components\label {table:variability} }
\tablehead{ 
\colhead{ }                       &   \multicolumn{3}{c}{ ------ Keck II  --------     }    & \multicolumn{3}{c} { ------  MIRI MRS ------ }      & \colhead{}     \\  [-10pt] 
 \colhead{}     &   \colhead{E}                 & \colhead{W } & \colhead{SW $+$ SE }     & \colhead{E} & \colhead{W}   & \colhead{SW  $+$ SE}  & \colhead{} \\  [-10pt]  
\colhead{}     &   \colhead{}                   &   \colhead{}    &  \colhead{Combined}       &\colhead{ }     & \colhead{}      &\colhead{Combined} &  \colhead{ }   \\ [-10pt]
\colhead{Wavelength}                           &   \multicolumn{6}{c}{------------------------------ Flux --------------------------------------- } &\colhead{} \\ [-10pt]
\colhead{(microns)}                        &  \colhead{(mJy)} & \colhead{(mJy)}  & \colhead{(mJy)}                                & \colhead{(mJy)} & \colhead{(mJy)}   & \colhead{(mJy)} & \colhead{}   \\ [-10pt]
}
\startdata
   7.9                              & 121             &  38.4                    & 123.0                        & 88       &35     &78.8     &               \\
 10.3                              &  72.6           &  49.6                    & 345.0                        & 76.9    &71.3  &178      &                \\
  10.8                             & 79.0            &  51.5                    & 281.0                        & 85       &73.4  & 249    &                   \\
 12.5                             &  86.8            & 44.3                     & 610.0                        & 85       &49.7  & 475     &                \\
 17.9                            &  78.0            & 93.9                     & 2720.0                       & 87      &117.5  & 1955.7  & Channel 3 LONG\tablenotemark{a} \\
 17.9                             & \ldots           & \ldots                   & \ldots                         & 86       &113    & 1822.5   & Channel 4 SHORT\tablenotemark{a}   \\
 20.8                             & 109.0         &117.0                     & 3700.0                       & 99.3    &154.3 & 2856.4  &                 \\
 24.5                             &  $<$155.0       & $ <$155.0                &6600.0                 &116.0    &241.6 & 4217.5   &       \\         
 \enddata
\tablenotetext{a}{For 17.9 $\mu$m, we list two MIRI MRS measurements since this wavelength is covered at the edge of  Channel 3 LONG and in Channel 4 SHORT. }
\end{deluxetable*}

The biggest surprise from the MIRI MRS continuum images presented in Figure \ref{fig:WL20continuum} is the discovery of a new member of the WL 20 system, WL 20SE,
adjacent to the previously known IRC source, which we are now calling WL 20SW.
The {\it continuum} mid-infrared appearance of the WL 20 system has previously been presented at 7.9, 10.3,  12.5, 17.9, 20.8, and 24.5 $\mu$m from Keck II imaging observations
with sufficient resolution at 7.9 and 10.3 $\mu$m to have resolved WL 20SE from WL 20SW  \citep{ResslerBarsony2001}.
The MIRI observations show WL 20SE to emit about 26.5\% of the flux of  WL 20SW at 5.3 $\mu$m, a wavelength at which both sources are well resolved by MIRI (see Figure 
\ref{fig:WL20continuum}). 
If the 5.3 $\mu$m flux ratio between WL 20SE and WL 20SW  were constant out to 7.9 $\mu$m, and the source fluxes had not varied between the time of the MIRI and the Keck II observations, 
the contribution of WL 20SE to the previously reported 7.9 $\mu$m flux would have been 33 mJy, a level near the detection limit, judging from the lowest level contour of 40 mJy
in the 7.9 $\mu$m plot of the WL 20 system (Figure 1 d) of \citet{ResslerBarsony2001}).

In fact, \citet{ResslerBarsony2001} did
report that WL 20S showed extended  structure, beyond the point-source appearances of WL 20E and WL 20W (see their Figure 7). Furthermore, it was noted that the extended structure
did not vary with wavelength, in contrast to other Class I objects, which generally exhibit increasing size with wavelength.

The combined fluxes from WL 20SW and WL 20SE had exhibited mid-infrared variability in the past on timescales of a few years \citep{ResslerBarsony2001}; thus, it is 
of interest to compare the newly acquired MIRI MRS flux measurements with previously published ones. These measurements are presented in
Table \ref{table:variability}, from which it is clear that the combined fluxes from WL 20SW and WL 20SE are consistently lower in 2023 
than they were in 1998.

\subsection{Gas Emission from Jets and Disk Winds}

The most spectacular discovery of the MIRI MRS observations of the WL 20 multiple system
is that of the parallel, twin, bipolar jets powered by WL 20SE and WL 20SW, in multiple transitions of 
[FeII] and [NiII], as well as in the higher-excitation [ArII] and [NeII] ionic lines 
(see Figures \ref{fig:FeII} $-$ \ref{fig:deconvolved} and Table \ref{table:results4}).

This is the first young system, lacking any associated, cold molecular outflow,
as traced by, for example, CO 1$-$0 or CO 2$-1$ emission, 
to exhibit ionized jets, discovered via centimeter radio observations \citep{Leous1991, Rodriguez2017}, but now resolved and imaged in the mid-infrared.
Recent JWST studies of Class 0 and Class I outflows point to a unified
picture of nested structures, the three outermost of which have wide opening angles, with the outermost layer (if detected) seen in scattered
near-infrared light, the next inner layer mapped in millimeter CO lines, within which the mid-infrared hydrogen emission is found.  
Collimated jets seen via their ionic emission lines are found along the symmetry axis of the wide-angled, outer layers
 \citep{Delabrosse2024, Federman2024, Narang2024, Tychoniecetal2024}.  In the case of WL 20,
the scattered light \citep{Ybarra2004} 
and CO outflow components are missing, since, at this evolutionary stage, there is no infall envelope left.

None of the detected molecular hydrogen
line maps of Figure \ref{fig:H2maps} show evidence of jets emanating directly from WL 20SE and WL 20SW.
Instead, the molecular hydrogen maps display a distinct biconical morphology with the apex coinciding with the
locations of WL 20SE and WL 20SW.  The ionized, parallel jets propagate along the symmetry axis of the molecular hydrogen cone.
At first glance, one might be tempted to interpret the H$_2$ structure as outlining biconical cavity walls. However,
examination of the gas distribution in Figure \ref{fig:MOM0} shows a lack of surrounding gas within which to form a cavity.
Rather, there is a clear spatial anticorrelation of the cold, CO molecular gas and the warm H$_2$ gas, strongly
supporting the interpretation that the H$_2$ structure originates from wide-angled disk winds.

These observations can be understood in the context of the coevolution of the protostellar envelope
and the jet sources it surrounds.  For the youngest, Class 0 protostars, the majority of the mass is
still in the envelope, and, as jets emerge from the central accreting sources, there is plenty of molecular material to entrain, 
resulting in molecular jets and outflows. As the sources evolve through the Class I stage, most of the mass is now 
concentrated in the protostars, with some remnant molecular infall envelope material, allowing for the appearance of molecular outflows,
wide-angled disk winds, and the mid-infrared ionized jets.  Finally, by the Class II stage, the molecular material from the original
protostellar envelope is gone (see Figures \ref{fig:MOM0} and \ref{fig:MOM1}), so, if enough disk accretion activity is still ongoing, the jets
will appear as purely ionic, since there is now no ambient molecular material to entrain.  Furthermore, this circumstance 
strengthens the case for the interpretation of the mid-infrared molecular hydrogen structures of Figure \ref{fig:H2maps} as resulting
from disk winds, which must be the origin of the observed H$_2$, since the envelope material is long gone.

\subsection {Disks in the WL 20 System}

\subsubsection{The Dust Disks Detected by ALMA}

Both the 1.3 mm and the 1.9 mm ALMA continuum observations, acquired at similarly high resolutions (0.14$^{\prime\prime}\ \times\ 0.11^{\prime\prime}$
and 0.096$^{\prime\prime}\, \times\, 0.16^{\prime\prime}$,  respectively) resolve the newly discovered WL 20SE source and the previously known WL 20SW into twin, edge-on disk structures
(see Figures \ref{fig:ALMA_Band4} and \ref{fig:MOM0}, respectively). 
The total 1.3 mm flux from the entire WL 20 system is 61.1 mJy (see Table \ref{table:results2}),
to be compared with the previously published 95 mJy acquired with the 12$^{\prime\prime}$ beam of the IRAM 30-meter, with a stated $\sim$ 20\% flux calibration uncertainty
 \citep{AndreMontmerle1994, MAN1998}.  Thus, within the stated uncertainties, 
  the high-resolution ALMA observations are consistent with having detected all of the continuum emission from the system.

The edge-on disk structures  resolved by ALMA explain both the Class I SED (spectral energy 
distribution) of WL 20SW (the dominant source at near- and mid-infrared wavelengths) and the additional $A_V =25$ inferred from its infrared spectrum, in addition to the $A_V = 16$ toward its northern neighbors.
These extinction values were previously inferred from near-infrared spectroscopy.
Photospheric spectral features evident in the near-infrared (2.06$-$2.49 $\mu$m) spectra of WL 20E and WL 20W were used to determine their 
spectral types and veilings \citep{BGB2002}. One can then apply various amounts of reddening, corresponding to known values of $A_V$, to match the observed spectral slopes
and fluxes to determine A$_V = 16$ for these sources.  Since the corresponding near-infrared spectrum of WL 20S (WL 20SW$+$WL 20SE) did not show any obvious 
absorption or emission features, its spectral slope and brightness were used to place constraints on its $A_V$ value. Under the assumption that its intrinsic spectrum
and K flux were the same as that of WL 20W, its spectrum could be matched with $A_V =  41$ \citep{BGB2002}.

The newly discovered edge-on disks of WL 20SE and WL 20SW by ALMA are well resolved, with diameters $\sim$ 100 au at 125 pc (see the right panel of Figure  \ref{fig:ALMA_Band4}).
The disks surrounding WL 20E and WL 20W, on the other hand, remain unresolved with ALMA, implying disk diameters less than 13 AU for each, for
an assumed distance of d$=$ 125 pc, derived from VLBA parallax measurements \citep{Loinardetal2008}.  This finding is perfectly in line 
with the results of the ODISEA (Ophiuchus Disk Survey Employing ALMA), which found the 1.3mm continuum disk sizes in Oph heavily weighted toward compact disks
with radii $<$ 15 AU for 85\% of detected objects \citep{Cieza_etal_2019}.  

Disk dust masses of just 3.3 M$_{\oplus}$ and 3.6 M$_{\oplus}$ are derived for WL 20E and WL 20W, respectively.
These values are in line with average disk dust masses derived for Class II objects in Corona Australis \citep{Cazzoletti_etal_2019},
 IC348 \citep{RR2018}, and Ophiuchus \citep{Cieza_etal_2019}.
By contrast, the disk masses around both WL 20SW and WL 20SE,  42 M$_\oplus$  and 24 M$_\oplus$, respectively (see Table \ref{table:results2}),  are in line with
the higher average disk dust masses derived for Taurus \citep{Andrews_etal_2013, Cox_etal_2017} and Lupus \citep{Ansdell_etal_2016}.

\subsubsection{X-ray excited 12.8$\mu$m [NeII] Disk Emission}

\begin{figure*}
\plotone{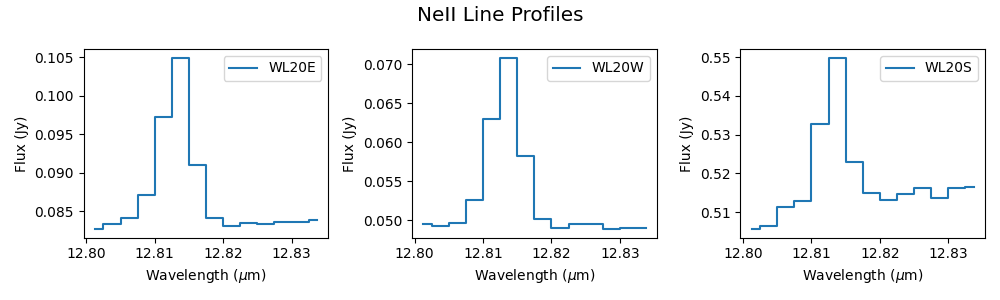}
\caption{ 
Line profiles of the 12.81 $\mu$m [NeII] emission toward each component of the WL 20 system.  At these wavelengths, WL 20SW is not resolved from 
its newly discovered companion, WL 20SE. Apertures used for the spectral extraction toward each source are the same as for the full spectra presented in Figure
\ref{fig:spectra} and centered on coordinates listed in Table \ref{table:results2}.
}
\label{fig:NeII}
\end{figure*}

The 12.81 $\mu$m [NeII] line was detected toward each of the WL 20 components, with the observed line profiles shown in Figure \ref{fig:NeII}.
Spectra were extracted through the same apertures as for the continuum flux measurements reported in Table \ref{table:results2}.
Continuum-subtracted [NeII] line fluxes, $A_{12.8}$, the extinction at 12.81 microns, and extinction-corrected NeII line luminosities are listed in Table \ref{table:NeII_lines}.
For reference, spectral types and inferred T$_{eff}$ are also listed \citep{BGB2002}.
The extinction at 12.81 $\mu$m was derived from the published values of $A_V$
of 16 for both WL 20E and WL 20W, and $A_V\ =$ 41 toward WL 20S \citep{BGB2002}.  We use $A_J\ =$ 0.282$A_V$ \citep{RL1985}, followed by the conversion
$A_{12.81}\ =\  0.16A_J$ for $R_V\ =\ 5.5$ for $\rho$ Oph \citep{WeingartnerDraine2001}. Taking into account the extinction toward each source, and using a distance
of 125 pc, we arrive at the intrinsic [NeII] line luminosities in the last column of Table \ref{table:NeII_lines}.

The WL 20 system was previously observed at 12.81 $\mu$m from the ground with VLT/VISIR through a 0.4$^{\prime\prime}$ slit at a resolution of R $=$ 30,000 corresponding 
to $\sim$ 10 km s$^{-1}$ \citep{Saccoetal2012}.
The observed coordinates were closest  to WL 20E and an upper limit of  $<$ 0.2 $\times$ 10$^{-14}$ erg cm$^{-2}$ s$^{-1}$ was established, quite close to our clear detection of the line toward
WL 20E at 0.24 $\times$ 10$^{-14}$ erg cm$^{-2}$ s$^{-1}$. 

The observed values for the [NeII] line fluxes agree nicely with predictions from X-ray excitation of neon in the warm upper atmospheres of disks around T Tauri stars, as first proposed
by \citet{Glassgold_2007}.   These authors calculated a 12.81 $\mu$m [NeII] flux of $\sim$ 10$^{-14}$ erg cm$^{-2}$  s$^{-1}$ for a fiducial T Tauri disk model from \citet{DAlessio_1999},
assuming a central star mass of 0.5 M$_{\odot}$, stellar radius of 2 R$_{\odot}$, T $=$ 4000K, $L_x \ = 10^{30}$ erg s$^{-1}$, and
accretion rate of 10$^{-8}$ M$_{\odot}$ yr$^{-1}$, for a face-on disk orientation at an assumed distance of 140 pc. 
           
We note that an X-ray flux of  $L_x \ = 1.16 \pm\ 0.09 \times\  10^{30}$ erg s$^{-1}$
from the DROXO (Deep Rho-Ophiuchi XMM-Newton Observation) survey is reported by \citet{Saccoetal2012} for the WL 20 system in its entirety, without
distinguishing among its individual components. 
The coordinates listed for the origin of  the X-ray emission in WL 20, (J2000) $\alpha\ =$ 16:27:15.9, $\delta\ =\ -$24:38:43.7 with a 1.1$^{\prime\prime}$ positional error,
originate from Table A.1. of \citet{Pillitterietal2010}. These coordinates, as previously noted, are closest to those of WL 20E.  Given that the FWHM of Newton/XMM is 
$\sim$6$^{\prime\prime}$, the deduced coordinates of the peak X-ray emission are biased toward the 
strongest emitter within the PSF, so we do not know the X-ray flux associated with each individual component of the WL 20 system. 

\begin{deluxetable*}{lccccc}
\tabletypesize{\scriptsize}
\tablecolumns{9}
\tablewidth{0pt}
\tablecaption{ [NeII] Line Fluxes and Additional Properties of the WL 20 System Components\label{table:NeII_lines} }
\tablenum{6} 
\tablehead{ 
\colhead{Source}  &\colhead{Spectral} & \colhead{T$_{eff}$} &\colhead{ $A_{12.8\mu m}$ }    & \colhead{Ne II Line}  & \colhead{Ne II Line}
\\[-10pt]  
\colhead{}             &\colhead{Type}       &\colhead{}              &\colhead{ }                                 & \colhead{Flux}          & \colhead{Luminosity}
\\[-10pt]  
\colhead{}            &\colhead{}               &\colhead{(K)}                  &\colhead{ }                                &\colhead{ (10$^{-14}$ erg cm$^{-2}$ s$^{-1}$)  } & \colhead{ (10$^{28}$ erg s$^{-1}$) }
 }
\startdata
   WL 20E           &    K7 IV/V                 &     4040                     &  0.72                                         &   0.24         & 0.87      \\
  WL 20W           &   M0 IV/V                 &     3800                      &  0.72                                        &    0.21        & 0.76      \\
  WL 20S            &    \ldots                    &     \ldots                      & 1.85                                         &   0.36         & 1.31      \\
  \enddata

\end{deluxetable*}

\subsubsection{ALMA Constraints on Disk Gas Masses}

We can estimate disk gas masses using the models of \citet{WilliamsBest2014},
with the measured  $^{13}$CO(2$-$1) and C$^{18}$O(2$-$1) line fluxes and flux upper limits
reported  in $\S$\ref{subsec:Band6} and a distance of 125 pc.  For the combined disks of WL 20SE and WL 20SW, these values lead to a 
combined gas mass of about 100 M$\oplus$ (3 $\times$ 10$^{-4}$ M$_{\odot}$) for an assumed CO/C$^{18}$O ratio of 550, or just $\sim$1 M$\oplus$ for the case of
a CO/C$^{18}$O ratio of 1650, meant to allow for selective photodissociation in the models. 
The combined gas mass of 100 M$\oplus$ leads to  a gas/dust ratio of just 1.5 for the combined disks of WL 20SE and WL 20SW,
and a factor of 100 lower for the case of the higher CO/C$^{18}$O ratio.

In stark contrast to the copious CO(2$-$1) gas emission toward the WL 20SW/WL 20SE system,
there is a striking lack of emission toward
either of the sources WL 20E or WL 20W (see left panel of Figure \ref{fig:MOM0}). 
Multiplying the 3$\sigma$ upper limits for the $^{13}$CO(2$-$1) and C$^{18}$O(2$-$1) line fluxes reported
in $\S$\ref{subsec:Band6} by 4 $\pi\ \times\ $ (d $=\  $125 pc)$^2$ for comparison with the grid of models presented in Figure 6 of  \citet{WilliamsBest2014},
we find that our observed upper limits fall well  below the lowest model gas mass of just 0.1 M$_{Jup}$ (32  M$_{\oplus}$). 
This {\it gas} mass upper limit is to be compared with
the derived dust masses of 3.3 M$_\oplus$ and 3.6 M$_\oplus$ for WL 20E and WL 20W, respectively.
One firm conclusion, as a result of this comparison, is
that the gas to dust ratio in these disks  is at least 10 times lower than the canonical ISM (Interstellar Medium) value of 100,
since otherwise, we would have 
had clear detections of both disks in each of the CO(2$-$1) isotopologues.

\subsection{Gas Emission from the Surroundings}\label{subsec:envelope}

\subsubsection{ALMA}

The twin disks of WL 20SW and WL 20SE lie near the peak of the flattened, elliptically shaped CO(2$-$1) structure evident in the left panel of Figure \ref{fig:MOM0},
and outlined by the gray ellipses in both Figure \ref{fig:MOM0} and Figure \ref{fig:MOM1}, which shows the Moment 1 maps in all three isotopologues, CO(2$-$1), $^{13}$CO(2$-$1), and C$^{18}$O(2$-$1).
A Moment 1 map is the integrated velocity-weighted intensity map divided by the integrated intensity map, and emphasizes the predominant
velocity distribution of the gas. A large-scale velocity gradient across this gaseous envelope
is most apparent in the CO(2-1) Moment 1 map in the left panel of Figure \ref{fig:MOM1}.  
Within the outlined elliptical area in this panel, the minimum and maximum velocities
are $-$1.41 km s$^{-1}$ and $+$3.62 km s$^{-1}$, respectively, relative to the cloud's $+$4 km s$^{-1}$ V$_{LSR}$.  

It is tempting to interpret the combined morphology and velocity distribution of the flattened $^{12}$CO(2$-$1) structure peaking 
on the twin disks of WL 20SE and WL 20SW as a pseudo-disk encompassing both of the smaller-scale twin disks.  
A pseudo-disk is simply a flattened  protostellar envelope first proposed as the natural consequence of including a uniform magnetic
field threading the density distribution of the singular isothermal sphere model of cloud core collapse \citep{GalliShu1993}.
If we were to interpret the elliptical CO(2$-$1) structure as a pseudo-disk, then for 
a central mass of 0.5 $-$ 1.0 M$_{\odot}$, the expected range of 
freefall velocities, $(2GM_*/R_{pseudo-disk})$, at a 310 AU radius (being the extent of the semi-major axis of the ellipse)
would be 1.1 km s$^{-1}$ to 2.4 km s$^{-1}$.
The expected velocity gradients
for infalling gas motions are, however, not observed in
the individual CO(2$-$1) velocity channel maps (not shown here).

Another possible interpretation of the observed velocity structure of the of the CO(2$-$1) gas within its emission peak is bulk rotation:
Assuming a central mass of 0.5 M$_{\odot}$ to 1 M$_{\odot}$, the corresponding range of 
Keplerian velocities expected at a radius of 310 AU
is 1.45 km s$^{-1}$ to 3.0 km s$^{-1}$, centered on the cloud's V$_{LSR}$ of 4 km s$^{-1}$.  Whereas the values of the observed velocity
extrema of $-$1.41 km s$^{-1}$ and $+$3.62 km s$^{-1}$ are close to those expected for a single, spatially resolved, Keplerian disk,
their locations are not -- they are not centered on WL 20SE and WL 20SW.
Close examination of the individual velocity channels reveals red- and blue-shifted gas components on each side of the twin disks,
consistent with the presence of two unresolved Keplerian disks blended together. Recall
that the large beam size of the CO(2$-$1) observations could not resolve the twin disks that 
were discovered via the much higher resolution continuum observations.

In the isotopologue maps of the central and right panels of Figure \ref{fig:MOM0},
we see progressively deeper into the envelope's gas structure, given the decreasing optical depths of the
$^{13}$CO(2$-$1) and C$^{18}$O(2$-$1) emission lines. Optical depth effects may account for the differing velocity structures encountered in the
corresponding Moment 1 maps.
The $^{13}$CO(2$-$1) Moment 1 map shows mostly red-shifted gas towards both disks, reaching
magnitudes of up to  $+$ 3 km s$^{-1}$ relative to the cloud's V$_{LSR}$.
In the  C$^{18}$O(2$-$1) Moment 1 map, the gas surrounding WL 20SW and WL 20SE is again red-shifted, but at lower velocities,
ranging from about 0.5 $-$ 1.5 km s$^{-1}$ relative to the V$_{LSR}$. 

An additional explanation for the gas structure in which the twin edge-on disks are embedded is that it represents
the leftover envelope gas which has mostly been obliterated by the twin bipolar jets emanating from WL 20SE and WL 20SW.
We can estimate the gas mass of this structure from the 
integrated intensity of the $^{13}$CO(2$-$1) emission
within the ellipse of the central panel of Figure \ref{fig:MOM0}, under the assumption of optically thin emission.
Adjusting the formula for the gas
mass estimate derived for $^{13}$CO(1$-$0) from
 \citet{SargentBeckwith1987} (following \citet{Scoville1986}) to $^{13}$CO(2$-$1):


\begin{eqnarray}
\label{eq:gasmass}
M(H_2)\,  =\, 1.37 \times 10^{-5}\, {\rm exp}(5.34/T_{ex})\,  (T_{ex} + 0.88)\,\times \nonumber \\
{\rm exp} \biggl(\frac{10.58}{ T_{ex} } \biggr) \, \biggl(\frac{D}{100 {\rm pc}} \biggr)^2\,  \biggl(\frac{10^{-6}}{X (^{13}CO)}\biggr)\, { \int {  S_{\nu}\, dv} } \ M_{\odot}  \nonumber \\
\end{eqnarray}

where $D$ is the distance to the source in units of 100 pc, $X(^{13}CO$) is the fractional abundance of $^{13}$CO with respect to H$_2$, and $T_{ex}$ is the excitation temperature of the gas.
Assuming $T_{ex}\ =$ 30K at 125 pc distance, and using the measured $S_{\nu}\, dv$ of 4.01 Jy km s$^{-1}$ in the ellipse in the central panel of Figure \ref{fig:MOM0}
yields a remnant envelope gas mass of just 4.5 $\times$ 10$^{-3}$ M$_{\odot}$.  By comparison, a core mass of 0.024 M$_{\odot}$  (or just 25 M$_{Jupiter}$) for WL 20 was
derived from
submillimeter continuum mapping of the $\rho$ Oph cloud at 850 $\mu$m with the 13$^{\prime\prime}$ beam
of the James Clerk Maxwell Telescope, adjusted for a distance of 125 pc \citep{Pattle2015}.

Placing this in an evolutionary context, it is useful to point out here that in a recent ALMA survey of 7 Class 0, 7 Class I, 
and 7 Flat Spectrum sources in Orion B, all sources exhibited beautiful red- and blue-shifted, bipolar CO (2$-$1) and 
$^{13}$CO (2$-$1)outflow structures \citep{hsiehetal2023}, 
in contrast with the case of WL 20S which completely lacks any trace of 
CO outflow activity (see left and middle panels of Figures \ref{fig:MOM0} and \ref{fig:MOM1}). 
Furthermore, in this same study, in which the
ambient cloud cores were mapped in  C$^{18}$ O(2$-$1), 
it was demonstrated that outflows remove a significant amount of gas from their parent cores. 
In the case of the WL 20 multiple system, it is amply evident that the mass already 
residing in the pre-main-sequence stars, exceeding 1 M$_{\odot}$ 
in just the WL 20E and WL 20W components alone, is far in excess of the ambient core gas mass, which is, at most, 0.024 M$_{\odot}$.

\subsubsection{Extended [NeII]}

\begin{figure*}
\plotone{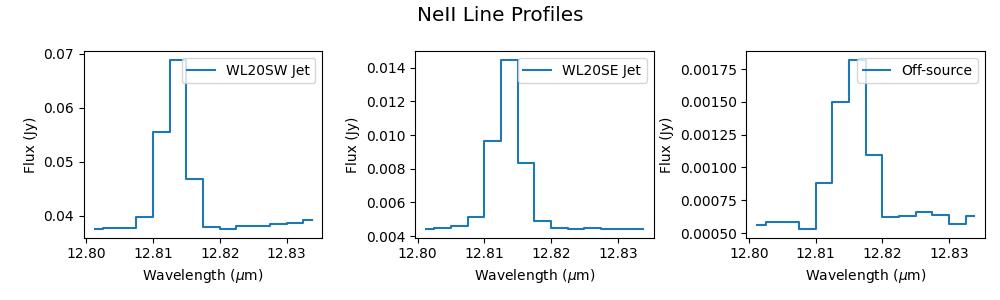}
\caption{Line profiles of the 12.81 $\mu$m [NeII] emission toward the WL 20SE and WL 20SW jet 
apertures outlined in Figure \ref{fig:deconvolved}. 
Note the difference in peak line fluxes. [NeII] emission in the off-source
aperture is a factor of $\approx$ 8 $-$ 40 below the emission found in the on-jet apertures.
}
\label{fig:NeII_jet_spectra}
\end{figure*}

\begin{figure*}
\plottwo{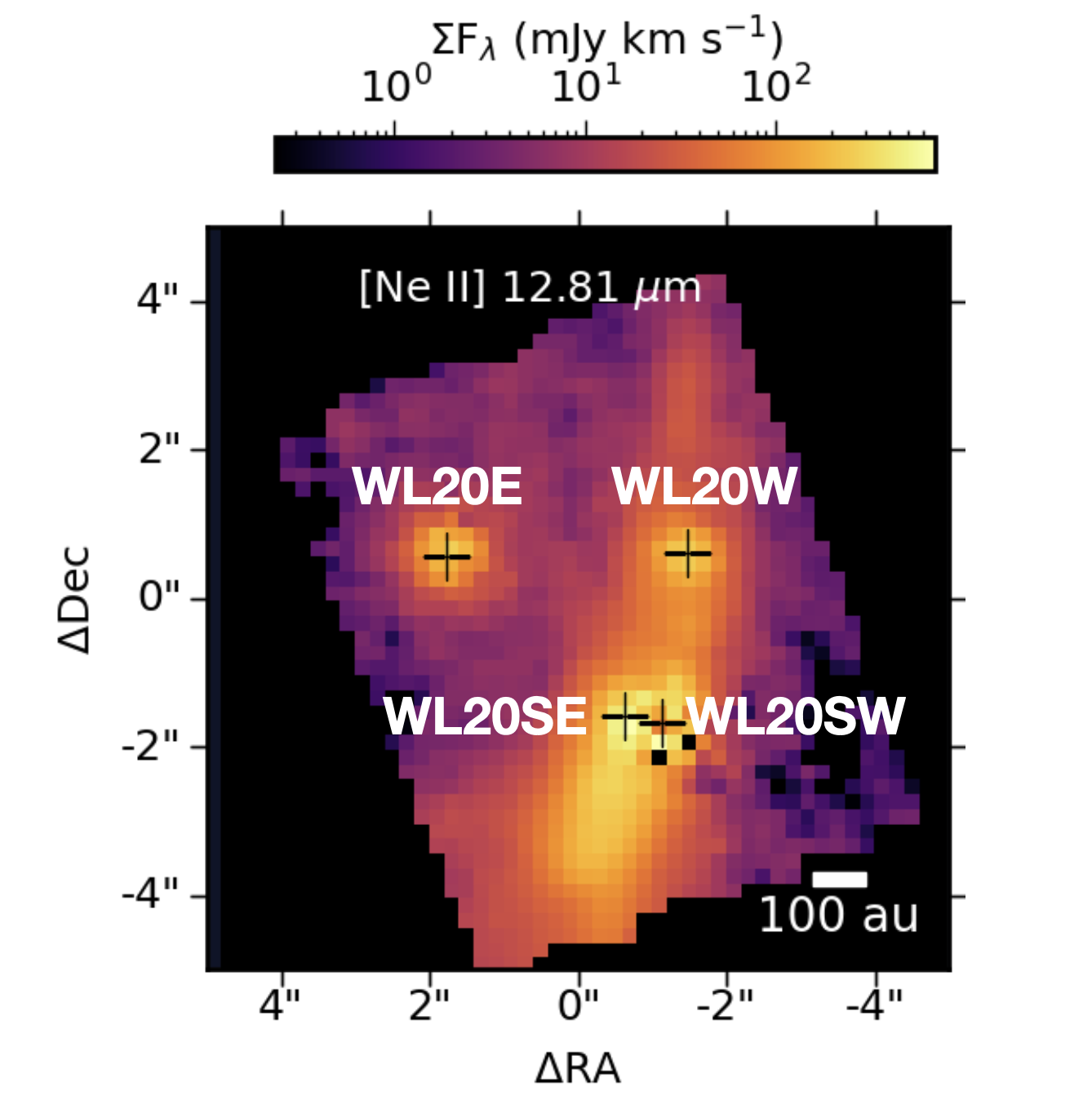}{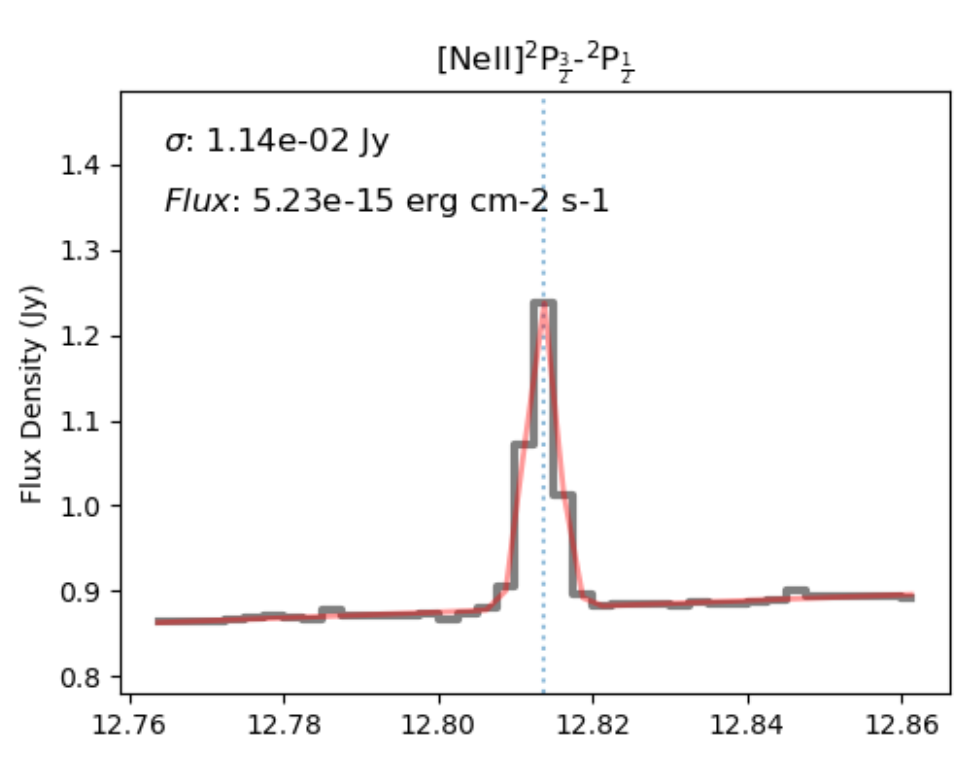} 
\caption{  {\bf Left panel}:  Continuum-subtracted 12.81 $\mu$m [NeII] emission in the
5.2$^{\prime\prime}\ \times 6.2^{\prime\prime}$ FOV.
Note the presence of diffuse, low-level [NeII] emission
throughout, in addition to emission from the twin jets, and from
localized [NeII] emission associated with each of the four disks in the system,
whose positions are indicated by black crosses and labelled.
A scale bar is displayed at bottom right and the color bar key to the flux scale is displayed at the top.
{\bf Right panel}:  The 12.81 $\mu$m [NeII] line profile, integrated over the entire MIRI MRS FOV, 
with the value of the measured, integrated line flux indicated.
}
\label{fig:NeII_line_map}
\end{figure*}

In Figure \ref{fig:NeII_jet_spectra}, 
we present the observed [NeII] line profiles
extracted through the jet apertures presented in Figure \ref{fig:deconvolved}. 
We also extracted a spectrum completely off-source to the northwest, through
a 1.00$^{\prime\prime}$ circular aperture centered at $\alpha_{2000 }\ =$16:27:15.703,
$\delta_{2000}\ =\ -$24:38:31.108,
displayed in the rightmost panel of Figure \ref{fig:NeII_jet_spectra}.
[NeII] emission was still detected, albeit with a peak flux $\sim$ 20$-$30 times weaker than that observed through the jet apertures.

This result led us to produce the continuum-subtracted, [NeII] integrated line map  in the left panel of Figure \ref{fig:NeII_line_map}, to examine the larger-scale spatial distribution
of the [NeII] emission.  Integrating the [NeII] line flux over the entire 5.5$^{\prime\prime}\ \times 6.2^{\prime\prime}$
FOV, shown in the right panel of Figure \ref{fig:NeII_line_map},
yields a value of 5.23 $\times$ 10$^{-15}$ erg cm$^{-2}$ s$^{-1}$, 
not corrected for extinction.  This value is to be compared with the 
6.28 $\pm$ 0.25 $\times$ 10$^{-14}$ erg cm$^{-2}$ s$^{-1}$ from the 
{\it Spitzer}/IRS c2d data  \citep{Saccoetal2012}.   
The {\it Spitzer} spectra were obtained  through a 4.7$^{\prime\prime}\ \times\ 11.1^{\prime\prime}$ aperture, with a resolution of $R\ =\ 600$.

As previously emphasized by \citet{Saccoetal2012}, the discrepancy between their VLT/Vizier [NeII] line flux upper limit and the 
value detected by {\it Spitzer} could be reconciled by the inferred presence of spatially extended
[NeII] emission originating from outflows. The new MIRI MRS imaging data confirm the shock-powered origin
of much of the observed [NeII] emission, in addition to the X-ray-excited [NeII] disk emission from each component of the WL 20 system, as 
can be seen in the left panel of Figure \ref{fig:NeII_line_map}. The [NeII] clearly fills the bipolar lobes traced by the H$_2$ maps of Figure \ref{fig:H2maps}.

\section{Putting It All Together: Conclusions and Summary}\label{subsec:synthesis}

\begin{figure*}
\begin{center}
\plotone{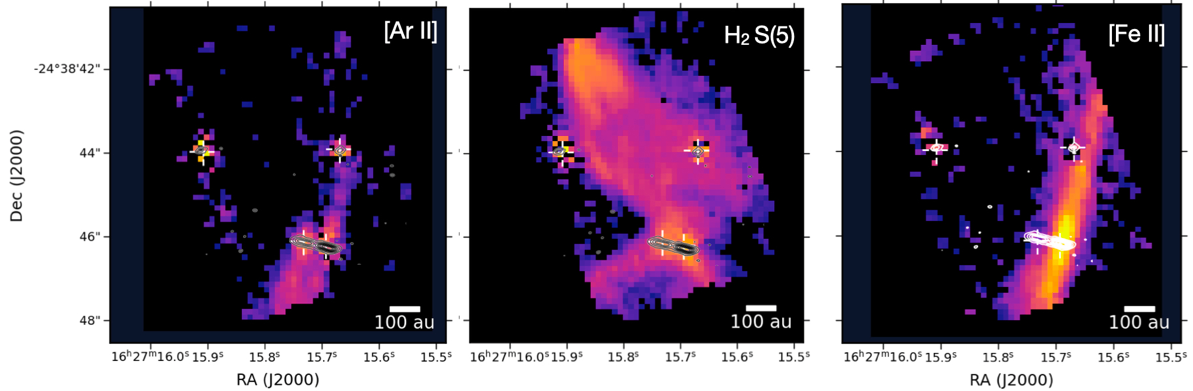}
\caption{Continuum-subtracted images of the [ArII] (left panel)  and [FeII] 5.34 $\mu$m (right panel) jets 
and the wide-angled, bipolar appearance of the H$_2$ emission (center panel), with the disks of all four sources, as imaged by ALMA
in the 1.9 mm continuum, superposed. A scale bar indicating the spatial scale  is drawn in the bottom right
of each panel.}
\label{fig:MIRIALMA}
\end{center} 
\end{figure*}

In Figure \ref{fig:MIRIALMA}, we present 
the combined 
discoveries of JWST MIRI MRS and ALMA in the WL 20 IRC system:
images of the twin disks, ionized twin jets, and the biconical molecular H$_2$ gas structure.

The four dust continuum disks detected by ALMA are represented by contours, and their locations are marked by the crosses
in each panel.
The highest-angular-resolution images showing the jets are shown in color in the left and right panels,
in the high-excitation [ArII] line at 6.985 $\mu$m,
and the low-excitation [FeII] line at 5.34 $\mu$m, respectively.
The jet driving sources, discovered with the MIRI MRS instrument in its highest angular resolution channels, coincide with the locations of the two small, edge-on disks, resolved by ALMA.

The molecular gas, in stark contrast with the ionized gas, is distributed in a double-cone-shaped structure, whose apex is centered 
on the two edge-on disks of WL 20SE and WL 20SW.
The double-cone-shaped morphology is shown in the middle panel of Figure \ref{fig:MIRIALMA},
which highlights the 
continuum-subtracted, H$_2$ 0-0 S(3) 9.67 $\mu$m emission.
The parallel jets propagate along the cone's central axis.
Portions of this bipolar conical gas structure clearly continue beyond the observed MIRI MRS field-of-view.

Based on near-infrared spectroscopy and ground-based, spatially resolved photometry,
the northern components, WL 20E and WL 20W, can be placed on model isochrones in
the Hertzsprung-Russell diagram.  Using this technique,
the most plausible system age is determined to be $\sim$ 2 $-$ 2.5 $\times$ 10$^{6}$ yr
\citep{ResslerBarsony2001}.
Given the proximity of the sources to each other, and taking into consideration the typical dimensions of pre-stellar cores, 
it is plausible to assume that the system evolved coevally,  leading to the conclusion that both WL 20SE and WL 20SW,
their disks, and their jets are also this old.
A later evolutionary stage for WL 20SE and WL 20SW is in keeping with the 
lack of any {\it molecular} outflows associated with these objects in the ALMA observations
of CO(2$-$1) and $^{13}$CO(2$-$1).
In addition, the structure of the relatively
optically thin C$^{18}$O(2$-1$) emission in the right panel of Figure \ref{fig:MOM0}
shows a pronounced anticorrelation
with the H$_2$ morphology of Figure \ref{fig:H2maps}, suggesting
that the combined outflows have dissipated
the original molecular gas core from which the system formed.
Such destruction of the molecular gas environment is also evident in the distribution
of [NeII] emission, which is not confined to the jets, but is distributed, albeit at a low level, 
throughout the MIRI MRS FOV (see Figures \ref{fig:NeII_jet_spectra} and \ref{fig:NeII_line_map}).
Finally, the core associated with the entire WL 20 system contains
only 0.024 M$_{\odot}$ of gas, accentuating the fact that the central objects
have already attained their birth masses.

The lack of a massive molecular core surrounding the system at this Class II evolutionary stage, 
simplifies the interpretation of the origin of the biconical H$_2$ gas -- it must originate in wide-angled disk winds, 
since there is no infalling or ambient cold gas to entrain. 
With regard to the presence of collimated, ionized jets, five have been found so far in Class 0 sources with MIRI MRS or NIRSpec
 \citep{Federman2024, Narang2024}, whereas one Class 0 source was found to be purely molecular 
\citep{rayetal2023, rayetalx2023}.
However, in a comprehensive K-band spectroscopic survey of 
26 Class 0 sources carried out with the MOSFIRE instrument at Keck I, 
whereas 90\% of sources showed H$_2$ emission, characteristic of shocks in outflows, 
only 20\% showed [FeII] emission lines, presumably
associated with narrow jets \citep{LeGouellec2024}.
This percentage of [FeII] detections in Class 0 sources may be a
lower limit, given the high line-of-sight extinction toward these most embedded protostars
and the fact that in many Class 0 sources in which the 5.34 $\mu$m [FeII] lines were detected by JWST,
the corresponding $H$ or $K$ band [FeII] lines were not.
For five Class I systems that have been 
investigated by JWST NIRSpec or MIRI MRS, all show the presence of ionized 
jets:  DGTauB \citep{Delabrosse2024}, the TMC 1 binary \citep{Tychoniecetal2024}.
and the HH46 IRS binary \citep{Nisini2024}. In a $K$-band spectroscopic survey carried
out with the Keck II telescope,
of 52 Class I and Flat Spectrum sources, 23 show H$_2$ emission and none were reported to show [FeII] \citep{Doppmann2005}. 

It has previously been suggested
that as the outflow evolves, the mass-loss rate decreases, velocities increase,
and the jet becomes progressively more ionized \citep{Nisini2015}.
In light of the information we have so far, it can be stated that the H$_2$ component
of the outflow activity, regardless of its origin, decreases in frequency with
evolutionary stage.

There are interesting differences amongst
the WL 20 system components worth pointing out as well: 
The WL 20SW/SE sources are both actively driving ionized jets, whereas neither WL 20E nor WL 20W are 
currently jet or outflow sources. The edge-on disks of WL 20SE and WL 20SW are well-resolved, both with 
extents of $\sim$ 100 AU, whereas the disks around WL 20E and WL 20W remain unresolved, suggesting
disk projected diameter upper limits of just 13 AU (see Figure \ref{fig:ALMA_Band4}).
Although small, the gas masses associated with the disks of WL 20SE and WL 20SW are measurable, whereas
any gas emission associated with the unresolved dust disks of WL 20E and WL 20W remains undetected, 
signaling a highly depleted gas to dust ratio compared with that of the ISM.

To summarize, combined JWST MIRI MRS and ALMA observations of the young, multiple, infrared companion system, WL 20, in the Ophiuchus 
star-forming region resulted in the discovery of the following:

\begin{enumerate}
\item A previously unknown companion to WL 20SW: WL 20SE
\item Twin, edge-on disks of $\sim$ 100 AU diameter,  with disk dust masses of  24 $\pm$ 4 M$_{\oplus}$
and 42 $\pm$ 2 M$_{\oplus}$ associated with WL 20SE and WL 20SW, respectively, and a combined {\it gas} mass of just
1 $-$ 100 M$_{\oplus}$
\item Unresolved disks with diameters $<$ 13 AU and dust masses of 3.3 $\pm$ 0.4 M$_{\oplus}$ and 3.6 $\pm$ 0.5 M$_{\oplus}$ 
for WL 20E and WL 20W, respectively, directly detected for the first time, with gas/dust ratios $\le$ 10
\item  Parallel, ionized jets, emanating from both WL 20SE and WL 20SW, seen in five transitions of [FeII],
two transitions of [NiII], and in [ArII] and [NeII]
\item The presence of extended, low-level [NeII] emission throughout
\item A biconical H$_2$ structure surrounding the ionized jets, observed in eight different mid-infrared H$_2$ lines, originating in wide-angled disk winds
\end{enumerate}

What is remarkable about these jets and the H$_2$ disk winds is the lack of any associated molecular line emission from cold gas  in 
the millimeter wavelength region.

\section{Acknowledgments}
M.B. would like to thank Ewine van Dishoeck for her leadership, enthusiasm, and encouragement during the preparation of this work;
the entire MIRI JOYS$+$ (JWST Observations of Young protoStars$+$) team. Diego Mardones for obtaining the Band 6 ALMA data, and Sue Terebey for 
fruitful discussions. We thank the anonymous referee for their attentive reading and numerous suggestions for improvement of the originally submitted manuscript.
The work of M.E.R. was carried out at the Jet Propulsion Laboratory, California Institute of Technology, under a contract
with the National Aeronautics and Space Administration. 
V.J.M.LG.'s research was supported by an appointment to the NASA Postdoctoral Program at the NASA Ames 
Research Center, administered by the Oak Ridge Associated Universities under contract with NASA.
M.L.v.G. acknowledges support from ERC Advanced grant 101019751 MOLDISK,
TOP-1 grant 614.001.751 from the Dutch Research Council (NWO) and The Netherlands
Research School for Astronomy (NOVA). Astrochemistry in Leiden is supported by the Netherlands Research School for Astronomy (NOVA).

This work is based on observations made with the NASA/ESA/CSA {\it James Webb Space Telescope}.
The JWST data presented in this article were obtained from the Mikulski Archive for Space Telescopes (MAST) at the Space Telescope Science Institute,
which is operated by the Association of Universities for Research in Astronomy, Inc., under NASA contract
NAS-5-03127 for JWST.  The specific observations analyzed can be accessed via 
\dataset[DOI:10..17909/d60d-mh65]{https://doi.org/10..17909/d60d-mh65} with the DataSet Title: ``MIRI/MRS WL 20'',
and \dataset[DOI:10.17909/pqce-5432]{https://doi.org/10.17909/pqce-5432} with the DataSet Title: ``MIRI/MRS Background used for WL 20 Data.'' 
The JWST MIRI data are from Program ID 01236, PI: Mike Ressler.

The following national and international funding agencies funded and supported the MIRI development:
NASA; ESA; Belgian Science Policy Office (BELSPO); 
Centre Nationale d'\'Etudes Spatiales (CNES); 
Danish National Space Center;
Deutsches Zentrum f\"ur Luft- und Raumfahrt (DLR); 
Enterprise Ireland;  
Ministerio de Economi\'a y Competividad;
The Netherlands Research School for Astronomy (NOVA);
The Netherlands Organization for Scientific Research (NWO); 
Science and Technology Facilities Council; 
Swiss Space Office; 
Swedish National Space Agency; and UK Space Agency.

This paper makes use of the following ALMA data: ADS/JAO.ALMA\#2019.1.01792.S and
ADS/JAO.ALMA\#2022.1.01734.S.
ALMA is a partnership of ESO (representing its member states), NSF (USA,) and 
NINS (Japan) together with NRC (Canada), MOST and ASIAA (Taiwan), and KASI (Republic of South Korea), in cooperation with the
Republic of Chile. The joint ALMA Observatory is operated by ESO, AUI/NRAO, and NAOJ.  This research has made
use of NASA's Astrophysics Data System Bibliographic Services, as well as the SIMBAD database,
operated at CDS, Strasbourg, France.  

{\it Software:} Numpy \citep{Harrisetal2020}; Astropy, a community-developed
core Python package for Astronomy \citep{Astropy2013, Astropy2018, Astropy2022}, Matplotlib \citep{Hunter2007}, and SuperMongo by Robert Lupton and Patricia Monger
({\tt https://www.astro.princeton.edu/~rhl/sm/ }).

\vspace{5mm}
\facilities{JWST(MIRI MRS), ALMA}





\bibliography{ms3sep2024}{}
\bibliographystyle{aasjournal}



\end{document}